\documentclass[useAMS,usenatbib]{mn2e}
\usepackage{times}
\usepackage{amssymb}
\usepackage{graphicx}
\usepackage{rotating}
\usepackage{lscape}
\usepackage{multirow}
\usepackage{fixltx2e}
\usepackage{footnote}

\def\totobs{795} 
\def\totres{584}
\def\totha{719}
\def\totdet{757}

\begin{document}

\title[The KMOS Redshift One Spectroscopic Survey]{The KMOS Redshift One Spectroscopic Survey (KROSS): Dynamical properties, gas and dark matter fractions of typical \boldmath$z\sim1$ star-forming galaxies} 

\author[J.P. Stott et al.]{John P. Stott$^{1, 2}$\thanks{E-mail: john.stott@physics.ox.ac.uk}, A. M. Swinbank$^{1,3}$, Helen L. Johnson$^3$, Alfie Tiley$^2$,  Georgios Magdis$^{2,4}$\newauthor  
Richard Bower$^{1,3}$, Andrew J. Bunker$^{2,5}$, Martin Bureau$^{2}$, Chris M. Harrison$^3$, Matt J. Jarvis$^{2,6}$ \newauthor
Ray Sharples$^{3,7}$, Ian Smail$^{3,1}$, David Sobral$^{8,9,10}$, Philip Best$^{11}$, Michele Cirasuolo$^{12}$ \\ 
$^{1}$ Institute for Computational Cosmology, Durham University, South Road, Durham, DH1 3LE, UK\\
$^{2}$ Sub-department of Astrophysics, Department of Physics, University of Oxford, Denys Wilkinson Building, Keble Road, Oxford OX1 3RH, UK\\
$^{3}$ Centre for Extragalactic Astronomy, Department of Physics, Durham University, South Road, Durham DH1 3LE, UK\\
$^{4}$ Institute for Astronomy, Astrophysics, Space Applications and Remote Sensing, National Observatory of Athens, GR-15236 Athens, Greece\\
$^{5}$ Affiliate Member, Kavli Institute for the Physics and Mathematics of the Universe, 5-1-5 Kashiwanoha, Kashiwa, 277-8583, Japan\\
$^{6}$ Department of Physics, University of the Western Cape, Bellville 7535, South Africa\\
$^{7}$ Centre for Advanced Instrumentation, Department of Physics, Durham University, South Road, Durham, DH1 3LE, UK\\
$^{8}$ Instituto de Astrof\'{\i}sica e Ci\^{e}ncias do Espa\c{c}o, Universidade de Lisboa, OAL, Tapada da Ajuda, 1349-018 Lisboa, Portugal\\
$^{9}$ Departamento de F\'{i}sica, Faculdade de Ci\^{e}ncias, Universidade de Lisboa, Edif\'{i}cio C8, Campo Grande, PT1749-016 Lisbon, Portugal\\
$^{10}$ Leiden Observatory, Leiden University, P.O. Box 9513, NL-2300 RA Leiden, The Netherlands\\
$^{11}$ SUPA, Institute for Astronomy, Royal Observatory of Edinburgh, Blackford Hill, Edinburgh, EH9 3HJ, UK\\
$^{12}$European Southern Observatory, Karl-Schwarzschild-Str. 2, 85748 Garching bei Muenchen, Germany\\
}

\date{}

\pagerange{\pageref{firstpage}--\pageref{lastpage}} \pubyear{2013}

\maketitle

\label{firstpage}

\begin{abstract}

The KMOS Redshift One Spectroscopic Survey (KROSS) is an ESO guaranteed time survey of \totobs\ typical star-forming galaxies in the redshift range $z=0.8-1.0$ with the KMOS instrument on the VLT. In this paper we present resolved kinematics and star formation rates for \totres\ $z\sim1$ galaxies. This constitutes the largest near-infrared Integral Field Unit survey of galaxies at $z\sim1$ to date. We demonstrate the success of our selection criteria with 90\% of our targets found to be $\rm H\alpha$ emitters, of which 81\% are spatially resolved. The fraction of the resolved KROSS sample with dynamics dominated by ordered rotation is found to be $83\pm5\%$. However, when compared with local samples these are turbulent discs with high gas to baryonic mass fractions, $\sim35\%$, and the majority are consistent with being marginally unstable (Toomre $Q\sim1$). There is no strong correlation between galaxy averaged velocity dispersion and the total star formation rate, suggesting that feedback from star formation is not the origin of the elevated turbulence. We postulate that it is {the} ubiquity of high  (likely molecular) gas fractions and the associated gravitational instabilities that drive the elevated star-formation rates in these typical $z\sim1$ galaxies, leading to the ten-fold enhanced star-formation rate density. Finally, by comparing the gas masses obtained from inverting the star-formation law with the dynamical and stellar masses, we infer an average dark matter to total mass fraction within $2.2\,r_e$ ($9.5\, \rm kpc$) of $65\pm12\%$, {in agreement with the results from hydrodynamic simulations of galaxy formation}.

\end{abstract}


\section{Introduction}
\label{sec:intro}

The star-formation rate density (SFRD) of the Universe peaks in the redshift range $z=1-3$ \citep{lilly1996,madau1996,hb2006,sobral2013}. At this epoch the average star formation rate (SFR) in galaxies was an order of magnitude higher than is observed locally. A major goal of galaxy evolution studies is to understand the conditions that occurred to enable this intense period of activity, during which the majority of the stars in the Universe were formed. 

Great advances have been made in charting the peak of star forming activity using multi-wavelength imaging and spectroscopy to obtain global properties of the galaxies at this epoch, e.g. the sizes, morphologies, SFRs, metallicities and gas content etc. of these galaxies (e.g. \citealt{doherty2004,doherty2006,erb2006,erb2006b,kassin2007,kassin2012,buitrago2008,dunne2009,peng2010,geach2011,vandokkum2011,bell2005,bell2012,stott2013,stott2013b,sobral2014}). 

To truly understand what drives this activity we need to resolve and study the processes that take place within the galaxies themselves. Instruments using Integral Field Units (IFUs), such as the Spectrograph for INtegral Field Observations in the Near Infrared (SINFONI) on the Very Large Telescope (VLT), provide spatially resolved spectroscopy of galaxies, with each spatial pixel (spaxel) having its own spectrum. This has been successfully employed by a number of groups to {resolve} relatively small samples of up to 100 galaxies on a time consuming one-by-one basis (e.g. \citealt{smith2004,genzel2006,genzel2008,shapiro2008,forster2009} [Spectroscopic Imaging survey in the Near-infrared with SINFONI, (SINS)]; \citealt{gnerucci2011,troncosco2014} [Assessing the Mass-Abundances redshift (Z) Evolution (AMAZE) and Lyman-break galaxies Stellar population and Dynamics (LSD)]; \citealt{queyrel2012,epinat2012,contini2012} [Mass Assembly Survey with SINFONI in VVDS (MASSIV)]; \citealt{swinbank2012,swinbank2012clump} [SINFONI observations of High-Z Emission Line Survey (SHiZELS) galaxies]; \citealt{law2009,wright2009}). {Other} studies have focussed on IFU observations of small samples of gravitationally lensed galaxies (e.g. \citealt{swinbank2006,jones2010tf,livermore2015}). By studying the kinematics and resolved star formation, surveys such as these have provided tantalising results on the internal processes of galaxies at $z\gtrsim1$.  However, although these samples cover a range in SFR and stellar mass, due to observational constraints (with the exception of the small numbers of lensed galaxies), the massive and highly star-forming galaxies {tend to be} over represented compared to the general star-forming population at a given redshift. 

Due to the limited sizes and potential biases of these $z\gtrsim1$ IFU studies it has been difficult to build a definitive picture of the internal properties of star-forming galaxies at this epoch. For example, the fraction of disc-like galaxies at these redshifts is found to be low (e.g. $\sim30\%$, \citealt{gnerucci2011}) but this may be because the highly star-forming galaxies probed in these surveys are more likely to be  disturbed than the typical population at that epoch \citep{stott2013}. It has also been demonstrated that these gaseous discs at $z\gtrsim1$ are highly turbulent but how this turbulence is maintained is unclear as {while \cite{green2014} and \cite{lehnert2013} suggest a strong link to star formation driven feedback, the results of \cite{genzel2011} imply that this is not the case}. The discs are also found to be marginally unstable to gravitational collapse with many dominated by clumpy star-forming regions \citep{forster2006,genzel2011,swinbank2012clump}. The Tully Fisher relation \citep{tf1977} is also found to evolve for galaxies at progressively high redshifts (although see \citealt{flores2006}), in that galaxies with the same rotation velocity in the distant Universe have lower stellar masses (e.g. \citealt{swinbank2006}). This suggests either higher gas fractions as measured directly by \cite{tacconi2010} or increased dark matter fractions or perhaps even both. Current observations of galaxies at $z\gtrsim1$ have therefore demonstrated that the internal properties of the galaxies at the peak epoch of star formation are indeed more extreme than their local counterparts but have we so far only been probing extreme examples? 

With the advent of near-infrared multi-object IFUs such as the $K-$band Multi Object Spectrograph (KMOS, \citealt{sharples2013}), it is now possible to study large well-selected samples of high redshift galaxies with much greater efficiency. KMOS allows for simultaneous observations with up to 24 IFUs within a 7.2 arcminute diameter radius and is thus perfectly suited to such a task. The KMOS Redshift One Spectroscopic Survey (KROSS) is a European Southern Observatory (ESO) guaranteed time survey undertaken by a team predominantly at Durham University and the University of Oxford, which has observed $\sim800$ mass-selected star-forming galaxies at $z\sim1$ (see also KMOS$\rm^{3D}$ [\citealt{kmos3d2014}] and the KMOS Science Verification programme KMOS-HiZELS [\citealt{sobral2013kmos,stott2014}]). With an order of magnitude increase in sample size compared to previous works, KROSS can study the resolved properties of galaxies in statistically significant sub-samples of parameter space, e.g. position within the SFR-stellar mass plane.

{To aid the comparison with theoretical models, the KROSS selection is kept as simple as possible and is dominated by galaxies on} and around the so called `main sequence' of star formation \citep{noeske2007,elbaz2011,karim2011}, the locus in SFR vs. stellar mass space occupied by the majority of star-forming galaxies at a given epoch. KROSS is designed to target the $\rm H\alpha$ emission line in these galaxies, which is an excellent tracer of ongoing star formation, less affected by dust obscuration than bluer indicators such as the UV continuum and [OII] emission line. We map the $\rm H\alpha$ and [NII] emission within the galaxies in order to measure the distribution of star formation and metallicity, the internal kinematics, and the role of any low-level active galactic nuclei (AGN) activity. KROSS is therefore the largest IFU-observed sample of typical star-forming galaxies at the closing stages of the peak in Universal star formation \citep{madau2014}. KROSS is also an excellent $z\sim1$ counterpart to the latest local IFU surveys such as the Calar Alto Legacy Integral Field Area Survey (CALIFA, \citealt{sanchez2012a}), Sydney-Australian-Astronomical-Observatory Multi-object Integral-Field Spectrograph survey (SAMI, \citealt{fogarty2012}) and Mapping Nearby Galaxies at Apache Point Observatory (MaNGA, \citealt{bundy2015}) which map $\rm H\alpha$ emission in galaxies at $z\sim0.1$.

In this paper we describe the KROSS sample, consisting of data taken in ESO periods 92, 93, 94 and 95 and use this to investigate the kinematic properties of the galaxies. We begin by discussing the sample selection, data reduction, and efficiency and then move on to kinematic modelling of the galaxies (\S\ref{sec:dat}). This yields the fraction of rotation and dispersion dominated galaxies and their dynamical masses (\S\ref{sec:dyn} and \S\ref{sec:mass}). We then compare the inferred dynamical masses to the stellar mass and the gas mass, from inverting the star formation law, to obtain gas and dark matter fractions (\S\ref{sec:fgasdm}). Finally, in \S\ref{sec:disc} we discuss the stability of the gaseous discs present in the $z\sim1$ star-forming population and give our conclusions in \S\ref{sec:sum}. Our results are compared to the output of the EAGLE hydrodynamic simulation.

We use a cosmology with $\Omega_{\Lambda}$\,=\,0.73, $\Omega_{m}$\,=\,0.27, and H$_{0}$\,=\,70\,km\,s$^{-1}$\,Mpc$^{-1}$. We note that $1''$ corresponds to 7.8\,kpc at $z=0.85$, the median redshift of the confirmed $\rm H\alpha$ emitting galaxies presented in this paper. All quoted magnitudes are on the AB system and we use a \cite{chabrier2003} IMF throughout.

\section{The Sample \& Data}
\label{sec:dat}
\subsection{Sample and target selection}

The KROSS survey is designed to study typical star-forming galaxies at $z=1$. We target these galaxies in the spectral range containing the redshifted $\rm H\alpha$ 6563\AA\ nebular emission line to obtain a measure of their ongoing star formation. The majority of the KROSS galaxies are selected to be those with known spectroscopic redshifts from various surveys, while the remainder ($\sim25\%$) are known $\rm H\alpha$ emitters from the HiZELS narrow-band survey \citep{sobral2013,sobral2015}. Table \ref{tab:fields} lists the observed fields{: UDS (UKIDSS [United Kingdom Infrared Telescope Deep Sky Survey] Ultra-Deep Survey); ECDFS (Extended {\it Chandra} Deep Field South); COSMOS (Cosmological Evolution Survey); and SA22,} and the spectroscopic and narrow-band surveys used {for each.}

The spectroscopic sample of galaxies in the redshift range $z=0.8-1.0$, places $\rm H\alpha$ in the $J-$band window that lies between two strong atmospheric absorption features. {These spectroscopic catalogues are: MUSYC (Multiwavelength Survey by Yale-Chile, \citealt{cardamone2010} and references therein), a sample of $\sim4,000$ galaxies from several surveys with $z_{AB}\lesssim24.5$ (for the largest sub-sample, \citealt{balestra2010}); UDS (\citealt{smail2008}, \citealt{bradshaw2013},  \citealt{mclure2013},, Akiyama et al., in prep and Simpson et al., in prep), a sample of $\gtrsim4,000$ galaxies, with $K_{AB}<24$ but with a significant AGN fraction which we remove with flagging (see \S\ref{sec:stat} for a discussion of the KROSS AGN fraction); VIPERS (VIMOS [VIsible MultiObject Spectrograph] Public Extragalactic Redshift Survey, \citealt{garilli2014} and \citealt{guzzo2014}), a survey of 100,000 galaxies with $I_{AB}<22.5$; VVDS (VIMOS VLT Deep Survey, \citealt{lefevre2005,lefevre2013} and \citealt{garilli2008}), a survey of 100,000 galaxies with $I_{AB}<22.5$, 50,000 with $I_{AB}<24$ and 1,000 with $I_{AB}<26$; and finally zCOSMOS (the spectroscopic component of COSMOS, \citealt{lilly2007}), which is a survey of 28,000 galaxies with $I_{AB}<22.5$.}

The majority of the sample {in the combined KROSS spectroscopic parent catalogue} are selected to be brighter than a magnitude limit of $K_{\rm AB}=22.5$. At redshift $z=0.8-1.0$ this $K-$band limit corresponds to a stellar mass limit of $\log (M_{\star} \rm [M_{\odot}])=9.3\pm0.5$ (see stellar mass calculation \S\ref{sec:msfr}). This limit was set by the feasibility simulations of the predicted sensitivity of KMOS we performed before the observing programme began, which assumed that the extended $\rm H\alpha$ emission followed the broad band galaxy photometry. We found that to guarantee resolved $\rm H\alpha$ in $\sim80\%$ of our galaxies in typical Paranal seeing conditions ($0.6''-0.8''$) and at our preferred on source integration time of 2-3 hours, we needed to select galaxies with seeing deconvolved broad-band half-light radii of at least $\sim 0.5''$ ($\sim4$ kpc), which on average corresponds to the magnitude/mass limit of $K_{\rm AB}=22.5$, $\log (M_{\star} \rm [M_{\odot}])=9.3\pm0.5$ \citep{stott2013}. {We note that other than this magnitude selection no formal size cut is applied to the sample.}

The HiZELS galaxies are drawn from \cite{sobral2013} and \cite{sobral2015}. These represent narrow-band $\rm H\alpha$ emitters at either $z=0.84$ (UDS and COSMOS fields) or $z=0.81$ (SA22 field, CF-HiZELS, \citealt{sobral2015}). The HiZELS galaxies are $\rm H\alpha$ emitters (typical flux, $F_{\rm H\alpha}>10^{-16} \rm erg\ cm^{-2}\ s^{-1}$) with AGN removed and are therefore likely to be extended \citep{sobral2013kmos,stott2014}. {The nature of this HiZELS selection means that these galaxies are SFR selected unlike the spectroscopic surveys which are magnitude, and therefore approximately stellar mass, limited. The wavelengths of the HiZELS narrow-band filters avoid regions of strong sky emission lines.} We note that $78\%$ of the HiZELS galaxies targeted are brighter than our {nominal} $K-$band limit ($K_{\rm AB}<22.5$)

To prepare the observations of the KROSS targets we assigned priority levels to the galaxies. As well as the magnitude/stellar mass criteria we include an $r-z$ colour diagnostic too, as these filters straddle the 4000\AA\ break {at the redshift of our galaxies and therefore provides} a good discriminant of red and blue galaxies. We assign the highest priority to those galaxies that are brighter than the $K-$band limit ($K_{\rm AB}<22.5$) and also bluer in $r-z$ colour than the typical $z\sim1$ `red sequence' of passive galaxies (i.e. $r-z<1.5$). Lower priorities are assigned to galaxies that are fainter than $K_{AB}=22.5$ and/or have a red colour (i.e. $r-z>1.5$). {Lower priority faint and red galaxies are still observed because it is not usually possible to fill all of the arms of a KMOS configuration with high priority galaxies, due to both target density and allowed arm positioning. }{The effect of down-weighting, although not exclusion of, the red galaxies from the sample will result in the selection of fewer passive galaxies and potentially fewer dusty starburst galaxies. It is difficult to observe $\rm H\alpha$ in both of these populations as the former have little or no ongoing star formation and the latter have a strong dust attenuation. The down-weighting of the passive galaxies will have no effect on our results as we primarily target star-forming galaxies. The down-weighting of the dusty starbursts may result in fewer disturbed, high specific star formation rate (sSFR) galaxies. However, these galaxies are rare so this is not a concern \citep{stott2013}}. 

{A final down-weighting of priority is applied to galaxies that would be strongly affected by night-sky emission lines. For this we compare the predicted observed $\rm H\alpha$ wavelengths from the galaxies' known redshifts to those of the night-sky emission line catalogue published in \cite{rousselot2000}. From our experience in \cite{stott2013b}, galaxies are given a lower priority if their redshifted $\rm H\alpha$ emission line is within $\rm 500\,km\,s^{-1}$ of strong OH lines, which we define as those with a relative flux greater than 50 (in the \cite{rousselot2000} catalogue).}

\begin{figure}
	\centering

\includegraphics[scale=0.5, trim=0 10 0 0, clip=true]{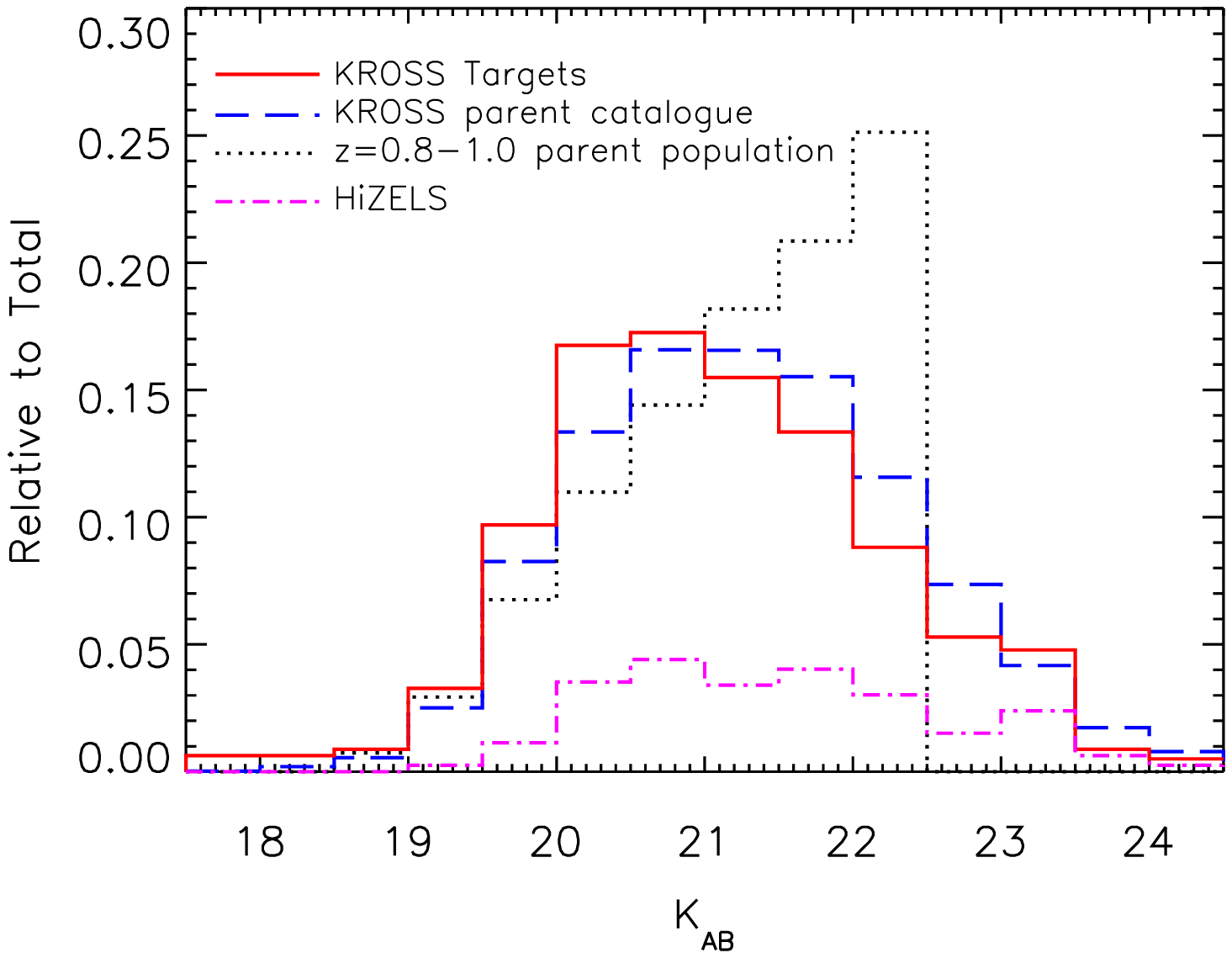}
		\includegraphics[scale=0.5, trim=0 0 0 25, clip=true]{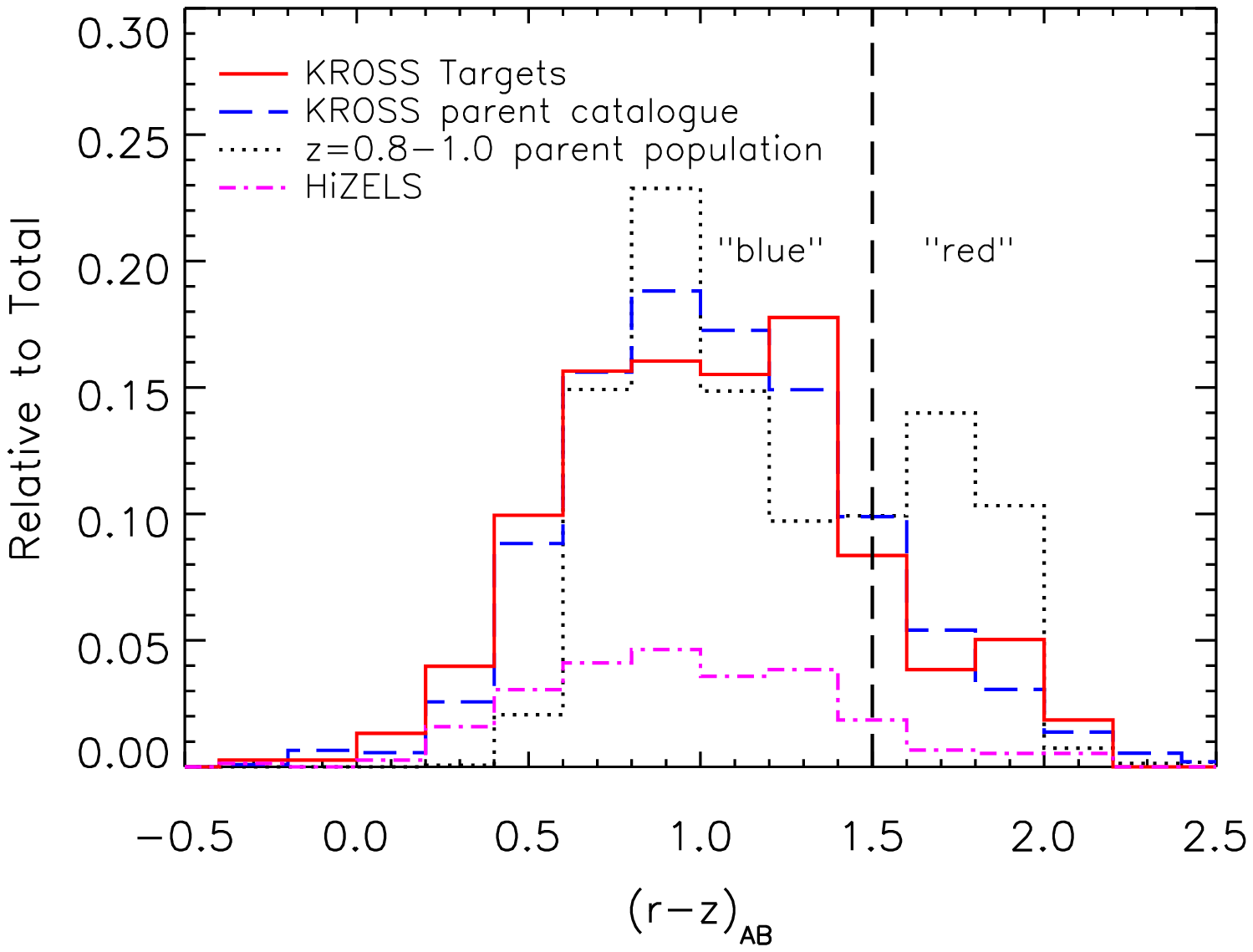}  
   	\caption[]{{\it Upper:} These are the $K-$band magnitude number counts for KROSS, its parent catalogue and an estimate of the parent population normalised to the number of galaxies in each. The parent population is a random sample of 3000 $K-$band selected galaxies sourced from the VIDEO survey (dotted, \citealt{jarvis2013}) with $K_{AB}<22.5$ and with a photometric redshift in the range $z=0.8-1.0$ to match KROSS. Also displayed is the combined KROSS parent catalogue of all galaxies with spectroscopic redshifts (plus HiZELS) within our survey fields and those that were actually targeted. The parent sample to KROSS is brighter than the underlying population, as you would expect, but the KROSS targets well sample this. The $K-$band magnitude distribution of the HiZELS galaxies only, is flatter than the parent sample and so they represent $43\%$ of the $K>22.5$ galaxies, reflecting their SFR selection. {\it Lower:} The distribution of $r-z$ colour. Again, the parent population is sourced from  the VIDEO survey (dotted) with $K<22.5$ and with a photometric redshift in the range $z=0.8-1.0$. This demonstrates that the parent sample to KROSS has a lower proportion of red galaxies than the underlying population, which is likely because it is easier to obtain spectroscopic redshifts for relatively dust-free star-forming, and therefore blue, emission line galaxies, rather than those that are passive or obscured. Again the KROSS targets well sample this.}
	\label{fig:samp}
\end{figure}

\begin{figure}
	\centering
	
		\includegraphics[scale=0.5, trim=0 0 0 0, clip=true]{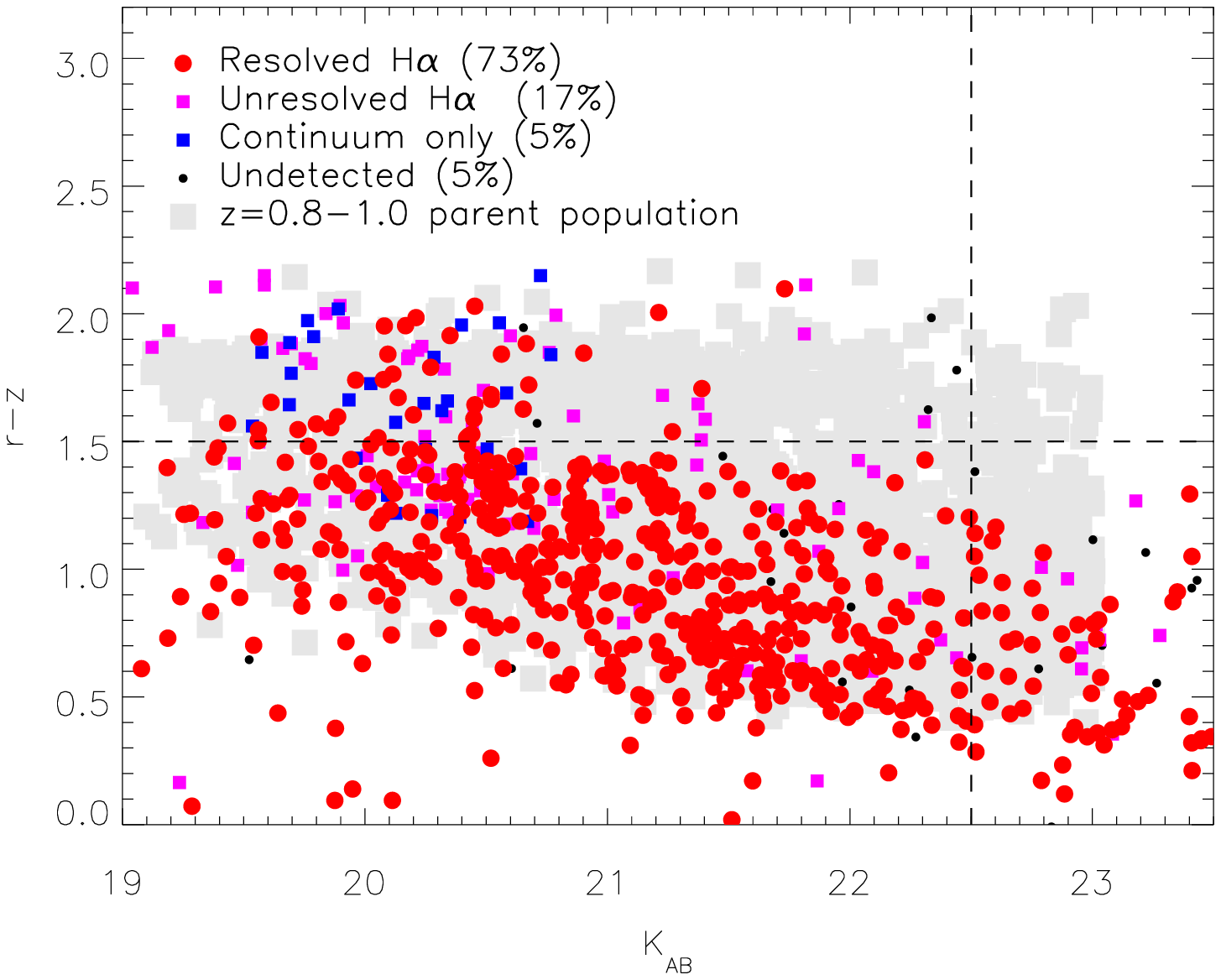}
			
	\caption[]{The $r-z$ vs. $K$ colour magnitude diagram for the KROSS sample. The red points are galaxies for which we resolved the $\rm H\alpha$ emission. The magenta squares are those galaxies in which we detect $\rm H\alpha$ ($>5\sigma$) but it is not spatially resolved and the blue squares are those where we detect continuum only. The small number of black points are those galaxies which are undetected in our spectra. The background region composed of light grey squares shows the position of the $z=0.8-1.0$ galaxy population from the VIDEO survey \citep{jarvis2013}, demonstrating that the KROSS galaxies are typical of the parent star-forming population.}
   	\label{fig:rzk}
\end{figure}

{In Fig. \ref{fig:samp} we plot the $K-$band number counts and the distribution of the $r-z$ colour for KROSS, its combined parent catalogue and an estimate of the parent population based on the VISTA (Visible and Infrared Survey Telescope for Astronomy) Deep Extragalactic Observations (VIDEO) photometric redshift catalogue \citep{jarvis2013}. This figure shows that KROSS is representative of its parent samples but brighter and with a lower proportion of red galaxies than the parent galaxy population, reflecting the priorities used for target selection.} {Fig. \ref{fig:samp} also highlights the colour and magnitude distributions of the HiZELS sample and demonstrates that they have a relatively flat distribution in $K_{\rm AB}$ and therefore represent a higher proportion ($43\%$) of the galaxies with $K_{\rm AB}>22.5$. This is because the $\rm H\alpha$ selection of HiZELS includes many low mass and therefore high sSFR galaxies. Fig. \ref{fig:rzk} is the $r-z$ versus $K$ colour magnitude diagram for the KROSS targets and displays the selection criteria.}

\begin{table*}
\begin{center}
\caption[]{A list of the extragalactic fields observed by KROSS and the parent catalogues from which we source our KMOS targets. N$_{\rm master}$ is the number of galaxies in the spectroscopic (and narrow-band) master catalogue. N$_{\rm obs}$ is the number of galaxies observed. The individual fields with their KROSS catalogue names are listed with their exposure times and average seeing.}
\label{tab:fields}
\small\begin{tabular}{llllllll}

\hline
Field &  $z$ Surveys&N$_{\rm master}$&N$_{\rm obs}$& Pointing &  Coordinates (J2000) &Exp. Time (s) & Seeing ('')\\
\hline

UDS& 1, 2, 3 & 7168 &209&&& \\
\hline
&&&&udsf2 &  02:17:24.5  -05:13:13  & 1800 & 0.7\\
&&&&udsf7 &   02:16:54.4 -04:59:04&    9000 & 0.5\\
&&&&udsf8 &    02:17:22.2 -05:01:06&   9000 & 0.9\\
&&&&udsf9 &     02:17:48.2 -05:01:02&  6000 & 0.6\\
&&&&udsf10 &  02:18:11.7 -05:00:04&     9000 & 0.7\\
&&&&udsf11 &    02:18:36.2 -05:01:49&   8400 & 1.2\\
&&&&udsf12 &  02:16:58.9 -04:46:42&     5400 & 0.7\\
&&&&udsf13 &   02:17:55.8 -04:43:00&    8400 & 0.5\\
&&&&udsf14 &     02:18:23.3 -04:46:12&  9000 & 0.8\\
&&&&udsf16 &  02:18:00.2 -04:53:43&     9000 & 0.5\\
&&&&udsf17 &  02:19:20.3 -04:51:58   & 9000 & 0.6\\
&&&&udsf18 &    02:19:26.2 -04:42:33 & 9000 & 0.9\\

\hline
ECDFS &  4, 5& 318&157& \\
\hline
&&&&ecdfsf0 &  03:32:06.4 -27:52:19  &  6000 & 0.7\\
&&&&ecdfsf1 &  03:32:28.4 -27:53:29 &  9600 & 0.7\\
&&&&ecdfsf1B &  03:32:28.4  -27:54:19 &    6000 & 0.4\\
&&&&ecdfsf2 &  03:32:53.5  -27:52:18  & 9000 & 0.5\\
&&&&ecdfsf2B & 03:32:53.7  -27:52:43  & 9000 & 0.6\\
&&&&ecdfsf3 & 03:32:06.6   -27:45:35 &  9000 & 0.6\\
&&&&ecdfsf3B & 03:32:07.4  -27:45:50  & 4800 & 0.8\\
&&&&ecdfsf4 & 03:32:32.5  -27:47:05 &   11400 & 0.9\\
&&&&ecdfsf6 & 03:32:07.9   -27:40:50 &    9000 & 0.5\\
&&&&ecdfsf7 &  03:32:38.4 -27:40:53 &   7800 & 1.0\\
\hline
COSMOS&  3, 6& 1743 &182& \\
\hline
&&&&cosmosf0 & 10:00:31.5 +02:13:34 &  9600 & 0.7\\
&&&&cosmosf1 &  10:01:06.1 +01:53:51 &  9000 & 0.5\\
&&&&cosmosf2 &  10:01:27.5 +01:57:26 &  9000 & 0.7\\
&&&&cosmosf4 &  09:59:25.3 +02:01:33 & 7200 & 0.6\\
&&&&cosmosf6 &  10:01:26.2 +02:03:15  & 8400 & 0.7\\
&&&&cosmosf9 &  10:00:48.9 +02:09:50 &  9000 & 0.5\\
&&&&cosmosf10 &  10:01:15.0 +02:09:38 &  6000 & 0.8\\
&&&&cosmosf12 &   10:01:01.2 +02:17:49 & 5400 & 0.9\\
&&&&cosmosf21 &09:59:43.7 +02:04:39     & 8400 & 0.8\\
&&&&cosmosf22 & 09:59:34.9 +02:16:43    & 9000 & 0.5\\

\hline
SA22&  2, 5, 7&  8283&247&\\
\hline
&&&&sa22f0 & 22:19:54.4 +01:10:16    &6000 & 0.8\\
&&&&sa22f5 &  22:20:21.3 +01:07:13 &  9000 & 1.0\\
&&&&sa22f6 & 22:18:46.0  +00:57:08  & 9000 & 0.6\\
&&&&sa22f8 &   22:11:59.9  +01:23:31 &9000 & 0.6\\
&&&&sa22f11 &  22:11:47.2 +01:14:32  &  9000 & 0.5\\
&&&&sa22f12 &  22:14:04.4 +00:55:55   &9000 & 0.5\\
&&&&sa22f13 &  22:18:21.8 +01:13:02 &  9000 & 0.8\\
&&&&sa22f16 &  22:20:39.8  +00:56:52 &  9600 & 1.0\\
&&&&sa22f19 &  22:20:33.7 +01:14:04 &  9000 & 1.0\\
&&&&sa22f21 & 22:13:03.9 +01:26:40  & 8400 & 0.7\\
&&&&SV1 &   22:19:30.3 +00:38:59&  4500 & 0.8\\
&&&&SV2 &    22:19:41.5 +00:23:20& 4500 & 0.6\\

\hline

\hline
\multicolumn{6}{l}{$^1$ UDS, \cite{bradshaw2013},  \cite{mclure2013},  \cite{smail2008}, Akiyama et al. (in prep) and Simpson et al. (in prep)}\\%, 
\multicolumn{6}{l}{$^2$ VIPERS, \cite{garilli2014} and \cite{guzzo2014}}\\
\multicolumn{6}{l}{$^3$ HiZELS, \cite{sobral2013}}\\
\multicolumn{6}{l}{$^4$ MUSYC, \cite{cardamone2010} and references therein}\\
\multicolumn{6}{l}{$^5$ VVDS, \cite{lefevre2005,lefevre2013} and \cite{garilli2008}}\\
\multicolumn{6}{l}{$^6$ zCOSMOS, \cite{lilly2007}}\\
\multicolumn{6}{l}{$^7$ CF-HiZELS, \cite{sobral2015}, \cite{sobral2013kmos} and \cite{stott2014}}\\

\end{tabular}

\end{center}
\end{table*}

\subsection{Observations and data reduction techniques}

The KMOS spectrograph consists of 24 integral field units (IFUs) that patrol a 7.2\,arcminute diameter field. Each IFU has an area of 2.8$''\times2.8''$ with $0.2''\times0.2''$ spatial pixels. At the average redshift of the KROSS targets ($z=0.85$) $2.8''$ corresponds to $\sim22\rm\ kpc$, which is well matched to mapping the KROSS galaxy properties out to several effective radii. {The full parent sample of potential KROSS targets consists of $\gtrsim17,000$ $z\sim1$ star-forming galaxies (see Table \ref{tab:fields}) but we choose to observe pointings with target densities similar to (or generally greater than) the density of IFUs within the KMOS field of view i.e. $>0.62$ galaxies per square arcminute. The observed pointings are those with the greatest number of high priority galaxies}. We note that the pointings are not over-dense structures of galaxies at the same redshift but are regions densely sampled by spectroscopic redshift surveys. {The number of galaxies satisfying the spectroscopic redshift cut for each field are given in Table \ref{tab:fields} along with the actual number observed and the details of the pointings.}

The KROSS observations were taken during ESO Periods 92, 93, 94 and 95 on the nights of: 22 November 2013; 7-8 and 24-26 December 2013; 21-22 February 2014; 19-27 August 2014; 13-15 and 29-31 October 2014; 25-28 January 2015; 10-12 April 2015; 4-7, 21-25 and 28-30 August 2015 (ESO programme IDs 092.B-0538, 093.B-0106, 094.B-0061 and 095.B-0035). We also include the KMOS-HiZELS Science Verification observations taken in 2013 on June 29, July 1 and September 25 (ESO programme ID 60.A-9460). For the full details of KMOS-HiZELS, which uses a similar selection criteria to KROSS but with a heavier weighting towards HiZELS sources, please see \cite{sobral2013kmos} and \cite{stott2014}. The median $J-$band seeing for the KROSS observations was $0.7''\pm0.2''$ with 70\% of the seeing conditions below $0.8''$ and $91\%$ below $1''$.  

We used the $YJ-$band grating in order to {observe} the H$\alpha$ emission, which at $z=0.8-1.0$ is redshifted to $1.181-1.312\mu$m. In this configuration, the spectral resolution is R\,=\,$\lambda$\,/\,$\Delta\lambda\,\sim\,3400$. We targeted up to 20 KROSS galaxies per pointing while deploying three IFUs (one per KMOS spectrograph) to (blank) sky positions to improve the sky-subtraction during the data reduction and typically one IFU to a star in order to monitor the PSF. Observations were carried out as three separate ESO observation blocks per pointing, with each using an ABAABAAB (A=object, B=sky) sequence, with 600s integration per position, in which we chopped by $>10''$ to sky, and each observation was dithered by up to $0.2''$. Therefore, the observations of one KROSS pointing took 4 hours of which the total on-source integration time was 2.5\,hrs per galaxy. 

To reduce the data, we used the {\sc esorex}\,/\,{\sc spark} pipeline \citep{davies2013}, which extracts the slices from each IFU then flatfields, illumination corrects and wavelength calibrates the data to form a datacube. However, we reduced each 600s frame separately (in order to have greater control of the reduction) and then removed the majority of the sky emission by combining individual AB (object-sky) pairs such that the sky frame is subtracted from the object frame. We then improved the sky OH subtraction in these individual 600s object `A-B' frames by subtracting the residual sky emission remaining in the A-B cube of the sky IFU, from the appropriate spectrograph. The best results were achieved by creating an average 1-D residual sky spectrum from an A-B sky IFU and then subtracting it (appropriately scaled) from each spaxel in the A-B object frame. This improvement is because the average residual sky does not add significant noise compared to subtracting the residual sky cube from the object cube on a spaxel-by-spaxel basis. 

We have found that at least two other variations of this residual sky subtraction technique are feasible and we list them here for those interested in reducing faint emission line sources with KMOS:

\begin{enumerate}
\item If no sky-IFUs were placed in the object frame (or the sky-IFU failed), it is possible to perform the residual sky subtraction by using one IFU per spectrograph for which there is no significant source flux detected (i.e. an IFU that is targeting a very faint source).  

\item Another successful technique is to median combine all of the other IFUs in the same spectrograph (i.e. all of the other object A-B frames) into one master sky residual frame which can then be averaged and subtracted as in (i), further reducing the noise added in the residual sky subtraction. However, this has the drawback that it is valid only for targets with no measurable continuum and at significantly different redshifts so that the spectral lines are averaged out of the final 1-D residual sky spectrum. It is however possible to run the reduction with a first pass, spatially and spectrally mask bright lines and continuum and then run the reduction again. 
\end{enumerate}

We suggest that any of the three techniques described above would be well suited to high redshift, faint emission line studies with KMOS, especially if no object frame sky-IFUs are available.

The $\rm (A-B)_{object}-(A-B)_{sky}$ frames are finally combined into a fully reduced datacube using a clipped average and then {oversampled} to $0.1''$ per spaxel. We flux calibrate our data using a standard star observed on one IFU per spectrograph {to ensure we have an independent zero point for each. A small sample of 10 galaxies were observed twice, in separate pointings. From the differences between the recovered $\rm H\alpha$ fluxes in these independent observations, we estimate the flux calibration error to be $\sim20\%$. }

The observations were spread over several different runs and semesters and so it was not always the case that all 24 arms were active, with up to four missing in the worst case. However, due to the flexibility of the large KROSS parent catalogue this had little impact on the survey.

\subsection{Sample statistics}
\label{sec:stat}

In this section we assess the efficiency of the KROSS selection technique. {First we calculate} the number of targeted KROSS galaxies compared to the number in which we detect resolved $\rm H\alpha$ (extent larger than the FWHM of the seeing disc), unresolved $\rm H\alpha$ ($>5\sigma$),  continuum only and finally those for which we detect no signal at all. This information is also displayed in Fig. \ref{fig:rzk} with the different coloured points denoting the different levels of detection. In terms of numbers, KROSS has observed \totobs\ galaxies \footnote{A catalogue of basic properties for the KROSS galaxies will be made available at the URL https://groups.physics.ox.ac.uk/KROSS/}. We detect $\rm H\alpha$ in \totha\ galaxies (90\%). The $\rm H\alpha$ emission is resolved in 81\% of these detections (\totres), which is an impressive return reflecting the success of our selection criteria and the simulations we performed before the observing programme began. Finally, if we include those for which we obtain significant continuum ({although none with obvious spectral features to obtain a redshift estimate}) the number is \totdet\ detections or 95\% of the KROSS sample. This leaves us with only 5\% non-detected galaxies which demonstrates the excellent efficiency of KROSS. 

For the priority 1 (P1) galaxies ($K_{\rm AB}<22.5$, $r-z<1.5$ or HiZELS $\rm H\alpha$ emitter) the recovery statistics are significantly improved. $\rm H\alpha$ is detected in 96\% of the P1 sample and resolved in 89\% of these, with $\rm H\alpha$ or continuum found in 97\%. {$\rm H\alpha$ is detected in 92\% of the HiZELS sample of which 83\% are spatially resolved. This is consistent with the entire KROSS sample. For the entire HiZELS sample there is only evidence that two galaxies have been assigned incorrect redshifts based on their narrow-band emission. These galaxies clearly have an [OIII]5007\AA\ emission in the narrow-band filter wavelength range as the KMOS spectra also show the [OIII]4959\AA\ and $\rm H\beta$ lines, meaning the galaxies are at $z=1.40$ and $z=1.42$ respectively and not $z=0.834$ and $z=0.847$. This is in agreement with the very low contamination rate estimated in \cite{sobral2013}. All of the HiZELS non-detections have HiZELS catalogue fluxes $F_{\rm H\alpha}<10^{-16} \rm erg\ cm^{-2}\ s^{-1}$ and so the likely reason for their non-detection is because they are faint. 

Of the non-HiZELS galaxies, 2.5\% have a measured redshift that has a larger than 10\% discrepancy with their catalogue redshift. We therefore estimate that a significant fraction of the 5\% non-detections may be explained by contamination. A further portion of this population will likely be continuum sources that are fainter than $K_{AB}=21$ (our approximate observed detection limit of continuum) and are therefore undetected in our relatively short exposures. Figures \ref{fig:samp} and \ref{fig:rzk} also indicate that the detection statistics are not uniform across the range of target selection criteria.  For example, the median colour of those with resolved $\rm H\alpha$ is $r-z=0.96$, whereas the galaxies with unresolved $\rm H\alpha$ are redder with a median $r-z=1.34$. The median colour of the galaxies with continuum but no emission lines is $r-z=1.65$ compared to those with resolved $\rm H\alpha$ and $K_{AB}<21$, which have a median $r-z=1.20$ (see Fig. \ref{fig:rzk}) }. This demonstrates that, as one may expect, it is the more massive, older, passive or dusty galaxies that show weaker $\rm H\alpha$ emission.

{Despite being careful to select against AGN in the parent spectroscopic samples, by removing galaxies flagged as AGN or X-ray sources, it is possible that there is some AGN contamination of KROSS. We can estimate the AGN fraction of KROSS by measuring the emission line ratio $\rm [NII]/H\alpha$. Using the $\rm [NII]/H\alpha$ diagnostics of \cite{kew2001} we find that none of the KROSS galaxies are identified as AGN as all have $\rm \log([NII]/H\alpha)<0.2$. There are four galaxies with $0.0<\rm \log([NII]/H\alpha)<0.2$ and a further 12 with $-0.2<\rm \log([NII]/H\alpha)<0.0$, which may indicate the presence of some mixed AGN/star-forming systems. Based on this line ratio diagnostic, the fraction of AGN in KROSS is likely $\ll5\%$. The $\rm [NII]/H\alpha$ line ratio can also be used to infer the metallicity of a galaxy, with a detailed analysis of resolved KROSS metallicities to be presented in Stott et al. (in prep.).}

\section{Analysis \& Results}
\label{sec:res}

\subsection{Stellar masses, star formation rates and spatial extent}
\label{sec:msfr}

{Since KROSS was carried out in some of the major extragalactic survey fields, there are extensive multiwavelength data allowing us to estimate stellar masses.} The stellar masses are derived by exploiting the multi-wavelength, optical -- {infrared ($U, g, B, V, R, I, z, J, K$ and IRAC: 3.6 and 4.5$\rm \mu m$) imaging { (\citealt{cirasuolo2007,lawrence2007,hambly2008,williams2009,cardamone2010,kim2011,muzzin2013,simpson2014,sobral2015} and references therein)}. The photometric bands were consistent between the fields except for SA22 where $g$ was used instead of $B$ and $V$ and no suitable IRAC imaging was available}. The mass estimates are obtained by fitting the spectral energy distributions (SEDs) of the galaxies in the KROSS parent catalogue using the {\sc hyperz} code \citep{hyperz2000} to compare the measured photometry with a suite of spectral templates from \cite{bc2003} {with the redshift fixed to that obtained from the KMOS $\rm H\alpha$ observation}. A full description can be found in Swinbank et al. (in prep.){, which provides a study of the strength of galaxy outflows and winds}. The median stellar mass of the entire observed KROSS sample is $\log(M_{\star} \rm [M_\odot])=10.0\pm0.1$. For those with resolved $\rm H\alpha$ this is also $10.0\pm0.1$ but for those detected in continuum only it is $10.9\pm0.1$. The median mass of those with resolved $\rm H\alpha$ that are brighter than our approximate continuum detection threshold ($K_{AB}=21$) is $\log(M_{\star} \rm [M_\odot])=10.4\pm0.1$, again demonstrating that the passive galaxies tend to be the most massive (see \S\ref{sec:stat}).

The SFRs of the galaxies are derived from the $\rm H\alpha$ emission line flux we measure for the galaxies. To obtain a `total' $\rm H\alpha$ flux for each galaxy we extract from the reduced datacube {an integrated} 1-D spectra in a $2.4''$ diameter circular aperture around the spatial $\rm H\alpha$ centroid. This centroid is found by collapsing the cube around a small wavelength range across the redshifted $\rm H\alpha$ line and fitting a 2-D Gaussian profile. {The diameter of the aperture corresponds to a large physical size of $\rm \sim18-19\,kpc$. For an exponential profile galaxy with the KROSS average half-light radius ($0.6''$, see below), $2.4''$ should contain 99\% of the flux, in the average seeing ($0.7''$)}. We then simultaneously fit the $\rm H\alpha$ and [NII] emission lines using Gaussian line profiles in order to extract their flux. In this fit we down-weight the spectra at the location of the OH skylines. 
 
The SFRs are calculated using the relation of \cite{kennicutt1998} assuming a \cite{chabrier2003} IMF. An individual extinction correction is applied to each galaxy based on the {stellar reddening} calculated from the SED fitting. This stellar based $A_{V}$ is converted to an extinction appropriate to the gas using the relation from \cite{wuyts2013},
 
 \begin{equation}
A_{V {\rm gas}} = A_{V\ {\rm SED}} (1.9-0.15\ A_{V\ {\rm SED}}),
\end{equation}

\begin{figure}
   \centering

\includegraphics[scale=0.5, trim=0 0 0 20, clip=true]{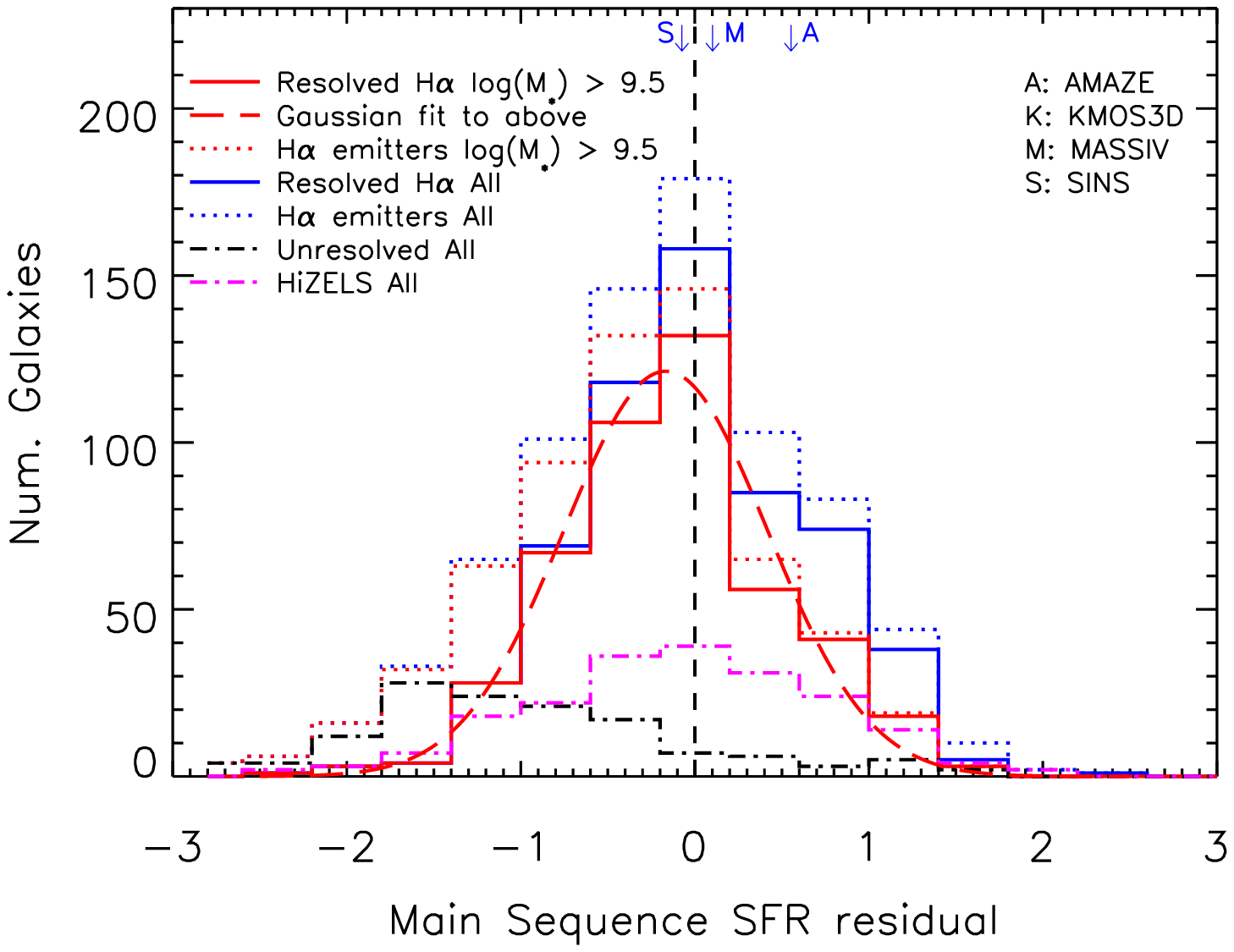}
\includegraphics[scale=0.5, trim=0 10 0 0, clip=true]{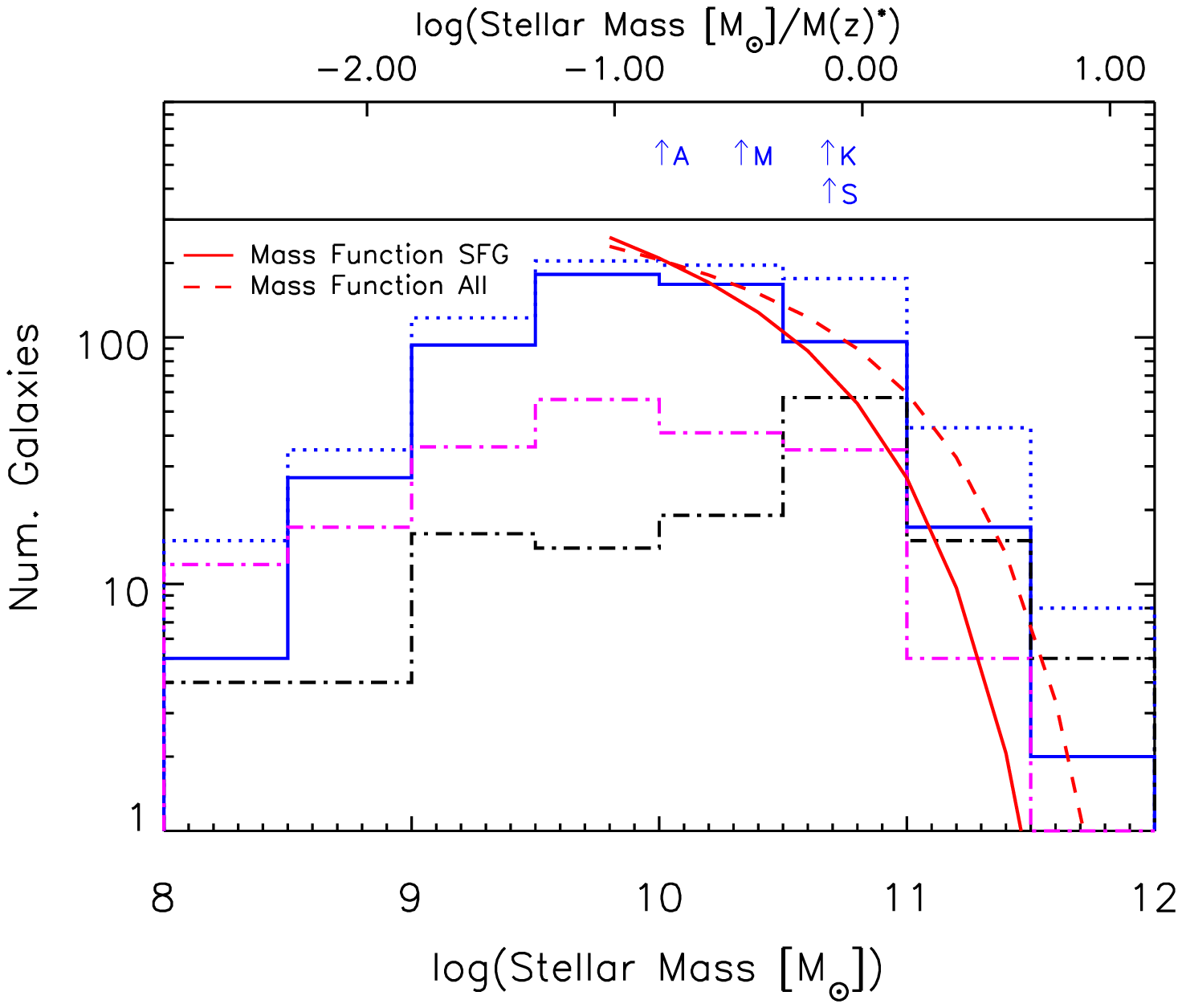}

\caption[]{{\it Upper:} The distribution of SFR of the KROSS sample relative to the $z=0.9$ main sequence of star formation from \cite{karim2011} represented by the vertical dashed line at 0\,dex. This demonstrates that the KROSS $\rm H\alpha$ emitters well describe the star-forming main sequence and are therefore typical of star-forming galaxies at this epoch. A distinction is made between the galaxies with mass above and below $\log(M_{\star} \rm [M_\odot])=9.5$ as that is the lower mass limit of the \cite{karim2011} main sequence. As one would expect, the unresolved sources are skewed towards the faintest $\rm H\alpha$ emitters in our sample. The positions of the median SFR of literature comparison samples relative to the main sequence at their median masses and respective redshifts are included for reference. {\it Lower:} The stellar mass function of KROSS and its sub-samples as above. The upper $x$-axis is relative to the Schechter function $M^{\star}$ from the star-forming galaxy mass function at $z=0.5-1.0$ \citep{muzzin2013b}. The normalised $z=0.5-1.0$ \cite{muzzin2013b} mass functions are included, confirming that KROSS well samples the galaxy population to $\log(M_{\star} \rm [M_\odot])\sim10$ ($\sim0.1M^{\star}$). It is clear that the unresolved sources are biased to high masses, indicating the generally more passive nature of massive galaxies. The positions of the median stellar mass of literature comparison samples relative to $M(z)^{\star}$ (the upper $x$-axis only) at their respective redshifts are included for reference (the KROSS median is $\log[M_{\star} \rm (M_\odot)]=10.0$). This demonstrates that the literature comparison samples tend to probe higher mass galaxies than KROSS, which better samples the sub-$M^{\star}$ population. }
   \label{fig:sfrhist}
\end{figure}

\noindent {which is consistent with the ratio of $\rm H\alpha$-to-far-infrared derived SFRs from a stacking analysis of {\it Herschel}/SPIRE observations of the KROSS galaxies (Swinbank et al., in prep.).} The median extinction-corrected SFR of the sample is $\rm 5\pm1 \ M_{\odot}yr^{-1}$ (Fig. \ref{fig:sfrhist}). 

{In Fig. \ref{fig:sfrhist} we show the distribution of SFR of the \totha\ $\rm H\alpha$ emitters in the KROSS sample about the $z=0.9$ main sequence from \cite{karim2011}. This plot demonstrates that the KROSS sample is indeed representative of galaxies on and around this trend in SFR with mass. A Gaussian fit to the distribution of the resolved $\rm H\alpha$ emitters with masses greater than $\log(M_{\star} \rm [M_\odot])=9.5$, has a peak at $-0.17\rm\,dex$ and a dispersion of $0.6\rm\,dex$. The width of this distribution is larger than that found by \cite{noeske2007}, who measure a dispersion of $0.35\rm\,dex$ for the main sequence at $z=0.85-1.1$ but we note that we are using $A_{V {\rm gas}}$, which introduces more scatter into the relation. If instead of individual extinction corrections, we use the average value $A_{V {\rm gas}}=1.43$ then we obtain a scatter around the main sequence of $0.4\rm\,dex$ in agreement with \cite{noeske2007}. Also shown in Fig. \ref{fig:sfrhist} is the distribution of the HiZELS galaxies only, which again peaks on the main sequence. {For a discussion of the properties of KROSS galaxies compared to their proximity to the main sequence, see \cite{magdis2016}}. Finally, Fig. \ref{fig:sfrhist} displays the distribution of the galaxies that could not be resolved in $\rm H\alpha$, which is strongly skewed to the lowest SFR galaxies as one would expect. The median SFR of these unresolved sources is $\rm 1.2\pm1.0 \,M_{\odot}  yr^{-1}$. The comparisons with the median SFR of other high redshift surveys, show that in general they also cluster around the main sequence, at their respective redshifts and median stellar masses, with only the AMAZE  \citep{gnerucci2011} galaxies at $z\sim3$ showing evidence of an elevated relative SFR.}

{The mass function of KROSS is also plotted in Fig. \ref{fig:sfrhist}, with comparison to the value of the Schechter  function $M^{\star}$ of the $z=0.5-1.0$ star-forming galaxy mass function of \cite{muzzin2013b}. This demonstrates that KROSS well samples the galaxy population to $\log(M_{\star} \rm [M_\odot])\sim10$ ($\sim0.1M^{\star}$). We can also see that the unresolved galaxies are skewed towards the high mass regime, which is a consequence of more massive galaxies having a tendency towards quiescence. The median stellar masses of the literature samples, compared to $M^{\star}$ at their respective redshifts, show that SINS \citep{forster2009} and KMOS$\rm^{3D}$ \citep{kmos3d2014} tend to probe more massive $\sim M^{\star}$ galaxies, while AMAZE probes a similar region of the mass function to KROSS, $\sim 0.1M^{\star}$.}

To measure the spatial extent of the star formation we calculate the effective radius of the $\rm H\alpha$. The half light radii ($r_e$) is calculated as the galactocentric radius {(centred on the peak of the $\rm H\alpha$ distribution)} at which the $\rm H \alpha$ flux in the 2-D collapsed continuum-subtracted cube is half of the total value. This is assessed non-parametrically and accounting for the apparent ellipticity and position angle of the galaxy we obtain from the disc model fitting (see \S\ref{sec:dyn}). {To correct for the effect of turbulence in the atmosphere, the seeing is subtracted in quadrature from the measured half-light radii}. The median {seeing-corrected} half-light radii of the $\rm H\alpha$ is found to be $4.3\pm 0.1 \rm \, kpc$, {with an $11\%$ median uncertainty on individual radii}.

\begin{figure*}
   \centering

\includegraphics[scale=1.05, trim=0 0 0 30, clip=true]{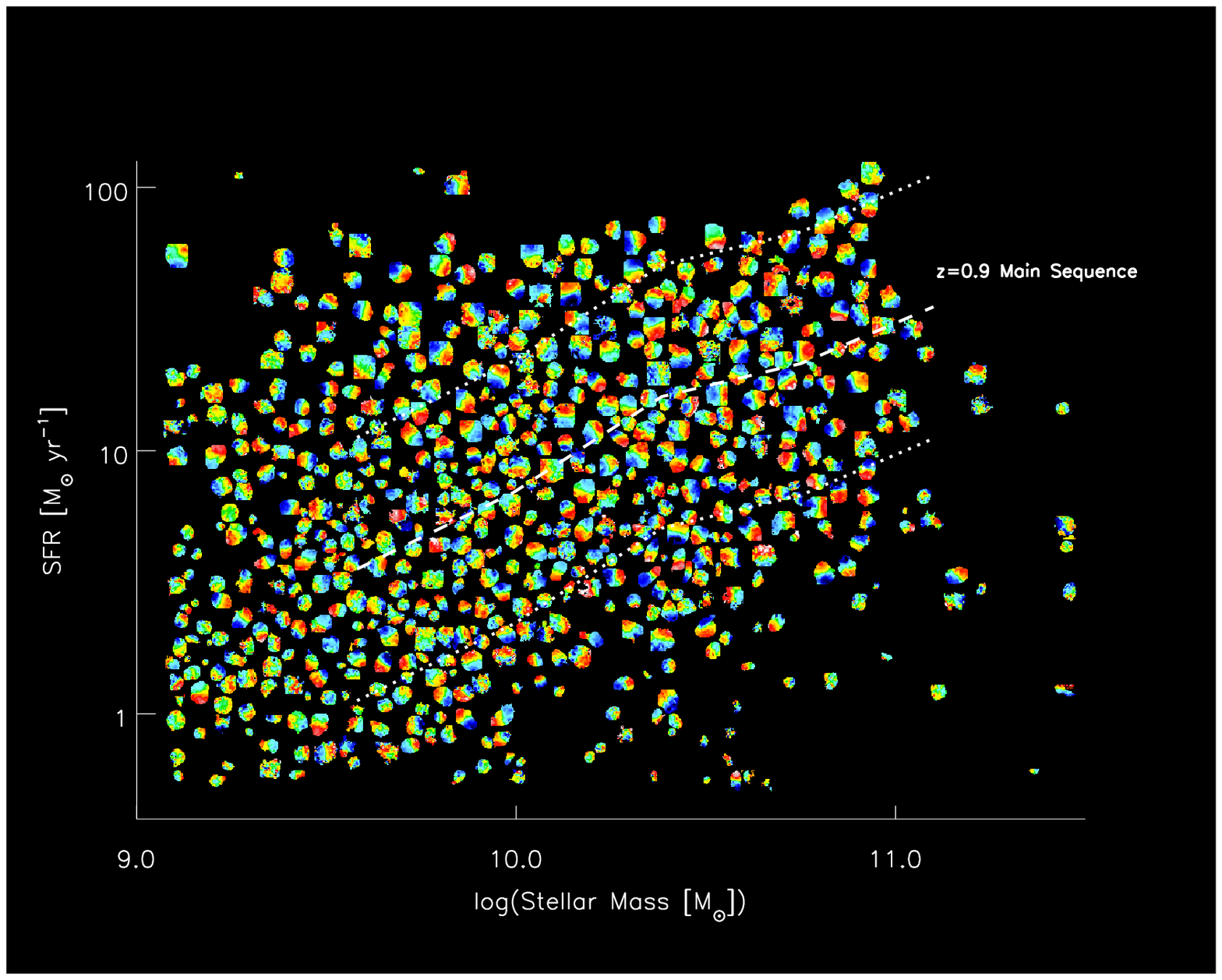}

\caption[]{The SFR plotted against stellar mass for the \totres\ resolved galaxies from the current KROSS sample with the data points represented by their velocity fields (normalised to their maximum observed velocities to make the rotation visible for a range of rotation speeds, $\sim30-300 \rm \ km\ s^{-1}$). For the velocity fields, red denotes a positive (recessional) velocity relative to the systemic redshift (green), while blue is negative. Note that the positions are approximate to avoid galaxy velocity fields from overlapping ({See Fig. \ref{fig:sfrhist} for the true distribution of galaxies in SFR and mass)}. The dashed line represents the location of the main sequence of star-forming galaxies at $z=0.8-1.0$ from \cite{karim2011} bounded by the dotted lines which represents a $\pm0.5$\,dex range. This demonstrates that our sample is typical of star-forming galaxies at this epoch and that the majority of them display ordered rotation.}
   \label{fig:ms}
\end{figure*}

{To further demonstrate how KROSS targets galaxies in and around the main sequence} and show that the majority display rotation (see \S\ref{sec:dyn}), in Fig. \ref{fig:ms} we plot up the SFR vs. stellar mass for those galaxies with resolved $\rm H\alpha$, with each galaxy represented by its velocity field. The velocity fields are normalised to their maximum value as it would be difficult to show the range of rotation velocities ($\sim30-300 \rm \ km\ s^{-1}$) with the same colour scheme. We over-plot the main sequence from \cite{karim2011}. {The galaxies are placed on the plot for clarity in such a way that they move to avoid overlapping with neighbours, however as was shown in Fig. \ref{fig:sfrhist}, it is clear that the KROSS sample is dominated by galaxies living on or around the main sequence.}

\subsection{Kinematics}
\label{sec:dyn}

In order to determine the kinematic properties of the KROSS sample we first need to create velocity maps for each galaxy with resolved $\rm H\alpha$ emission. To measure the spatially resolved gas velocities of each galaxy, we fit the H$\alpha$ and [NII] emission lines in each spaxel using a $\chi^{2}$ minimisation procedure (accounting for the increased noise at the positions of the sky emission lines). We initially try to fit to the H$\alpha$ line in one $0.1''$ spaxel but if this fit fails to detect the line with a signal-to-noise $>$\,5, the region is increased to a 0.3$''$\,$\times$\,0.3$''$ region taking in surrounding spaxels. If the S/N$>$\,5 criterion is still not met then the region is expanded to 0.5$''$\,$\times$\,0.5$''$ and finally 0.7$''$\,$\times$\,0.7$''$ (corresponding to the FWHM of the average seeing). The H$\alpha$ and [NII] emission lines are fitted allowing the centroid, intensity and width of the Gaussian profile to find their optimum values (the central wavelength and FWHM of the H$\alpha$ and [NII] lines are coupled in the fit). Uncertainties are then calculated by perturbing each parameter, one at a time, allowing the remaining parameters to find their optimum values, until $\chi^2$ is minimised. 

The method above produces flux, central line wavelength and linewidth maps of H$\alpha$ and [NII]. The total H$\alpha$ flux map is in excellent agreement with the `total' H$\alpha$ fluxes obtained from the $2.4''$ aperture we apply to the datacube. These flux maps can then be converted to a map of SFR using the \cite{kennicutt1998} relation assuming a \cite{chabrier2003} IMF and the $A_{V {\rm gas}}$ value calculated in \S\ref{sec:msfr}. The velocity map is created by taking the wavelength of the H$\alpha$ line in each spaxel and converting this to a velocity relative to the systemic redshift (as measured from the galaxy integrated 1-D spectrum).

Many of the KROSS galaxies have H$\alpha$ velocity fields which resemble rotating systems (characteristic `spider' patterns in the velocity fields and line of sight velocity dispersion profiles which peak near the central regions) as shown in  Fig. \ref{fig:ms}. Therefore, we attempt to model the two dimensional velocity field in order to identify the kinematic centre and major axis. This simple model is fitted to all spaxels with a S/N$>$\,5, all of which are assumed to be independent even if they are the result of binning their surrounding pixels. We follow \cite{swinbank2012} to construct two dimensional models with an input rotation curve following an arctangent function 

\begin{equation}
v(r)\,=\,\frac{2}{\pi}\,v_{\rm asym}\,{\rm arctan}(r/r_t), 
\label{eq:arc}
\end{equation}

\noindent where $v_{\rm asym}$ is the asymptotic rotational velocity and $r_{\rm t}$ is the effective radius at which the rotation curve turns over \citep{courteau1997}.  The suite of two dimensional models which we fit to the data have six free parameters ([x,y] centre, position angle (PA), $r_{\rm t}$, $v_{\rm asym}$, and disc inclination) and we use a genetic algorithm \citep{charbonneau1995}, {convolved with the PSF at each iteration, { assuming a flat $\rm H\alpha$ flux distribution}} to find the best-fit model \citep[see][]{swinbank2012}. The uncertainties on the final parameters are estimated to be the range of parameter values from all acceptable models that fall within a $\Delta\chi^2=1$ of the best-fit model. These uncertainties are carried through the following analyses. This is important as parameters such as the inclination, which can be difficult to assess from the kinematics, may have a strong impact on the determination of the rotation velocity, particularly at low inclinations. The median uncertainty of the inclination angle is found to be $14\%$, {although we note that this uncertainty is model dependent and may not fully reflect the robustness of this parameter. In Harrison et al. (in prep.), a KROSS study of specific angular momentum, we will also derive stellar disc inclinations for a sub-sample of our galaxies with {\it Hubble Space Telescope} imaging.} 

\begin{figure}
   \centering

\includegraphics[scale=0.5, trim=0 0 0 0, clip=true]{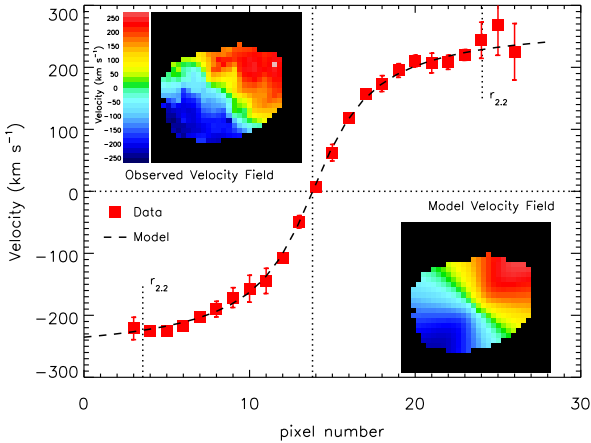}

\caption[]{An example of the observed and model velocity maps with the extracted 1-D rotation curve and an arctangent fit for a KROSS galaxy at $z=0.876$ in the COSMOS field. This demonstrates that $r_{2.2}$ probes the relatively flat region beyond the turnover in the rotation curve. All of the KROSS rotation curves are plotted in Tiley et al. (in prep.), which is a detailed analysis of the KROSS Tully Fisher relation. }

   \label{fig:vcurve}
\end{figure}

The best fit kinematic model produces a kinematic centre and position angle of the disc allowing us to extract the one dimensional rotation curve and velocity dispersion profiles from the major kinematic axis of each galaxy. {We note that the scatter between the kinematic centroid and the $\rm H\alpha$ centroid, calculated in \S\ref{sec:msfr}, is $0.19''$ (1.5\,kpc) and that the choice of $\rm H\alpha$ or kinematic centroid for the analysis that follows has no effect on our conclusions.} The majority of the KROSS galaxies possess clear rotation curves which turn over or flatten (see also \citealt{sobral2013kmos} and \citealt{stott2014}). {An example rotation curve (not corrected for beam smearing) is plotted in Fig. \ref{fig:vcurve}. All of the KROSS rotation curves are plotted in Tiley et al. (in prep.), which is a detailed analysis of the KROSS Tully Fisher relation.}

\begin{figure*}
   \centering

\includegraphics[scale=0.37, trim=20 0 10 0, clip=true]{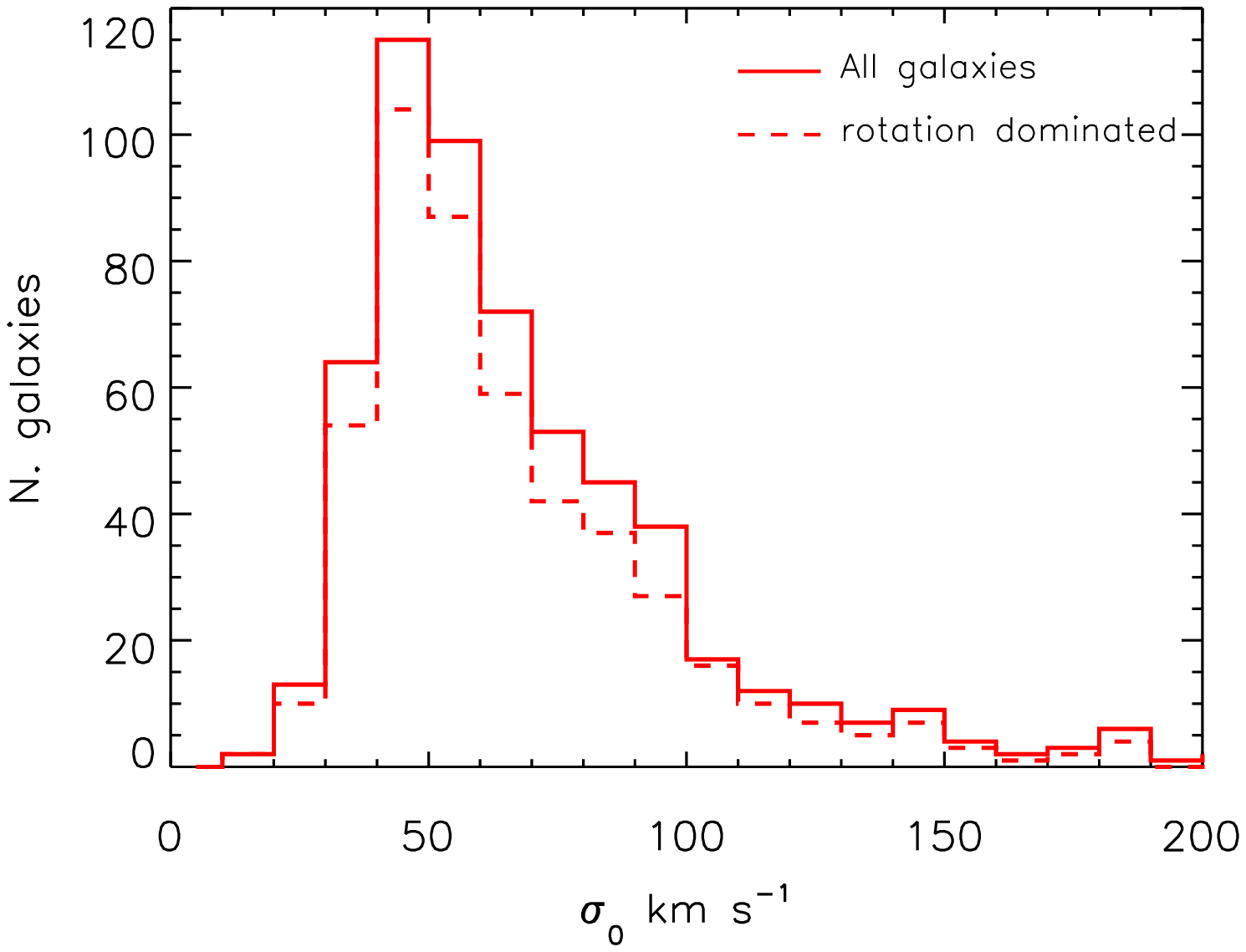}\includegraphics[scale=0.37, trim=20 0 10 0, clip=true]{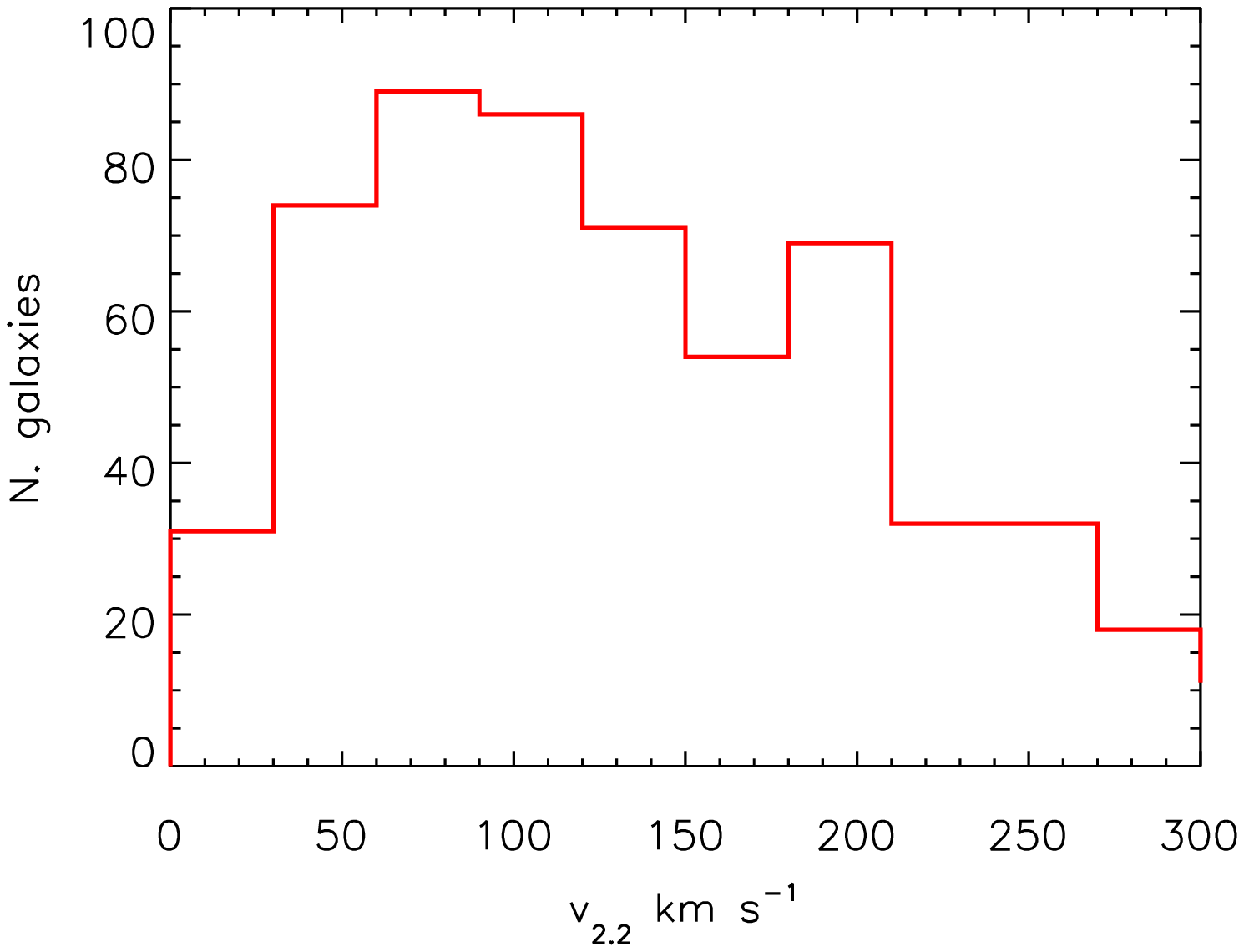} \includegraphics[scale=0.37, trim=20 0 10 0, clip=true]{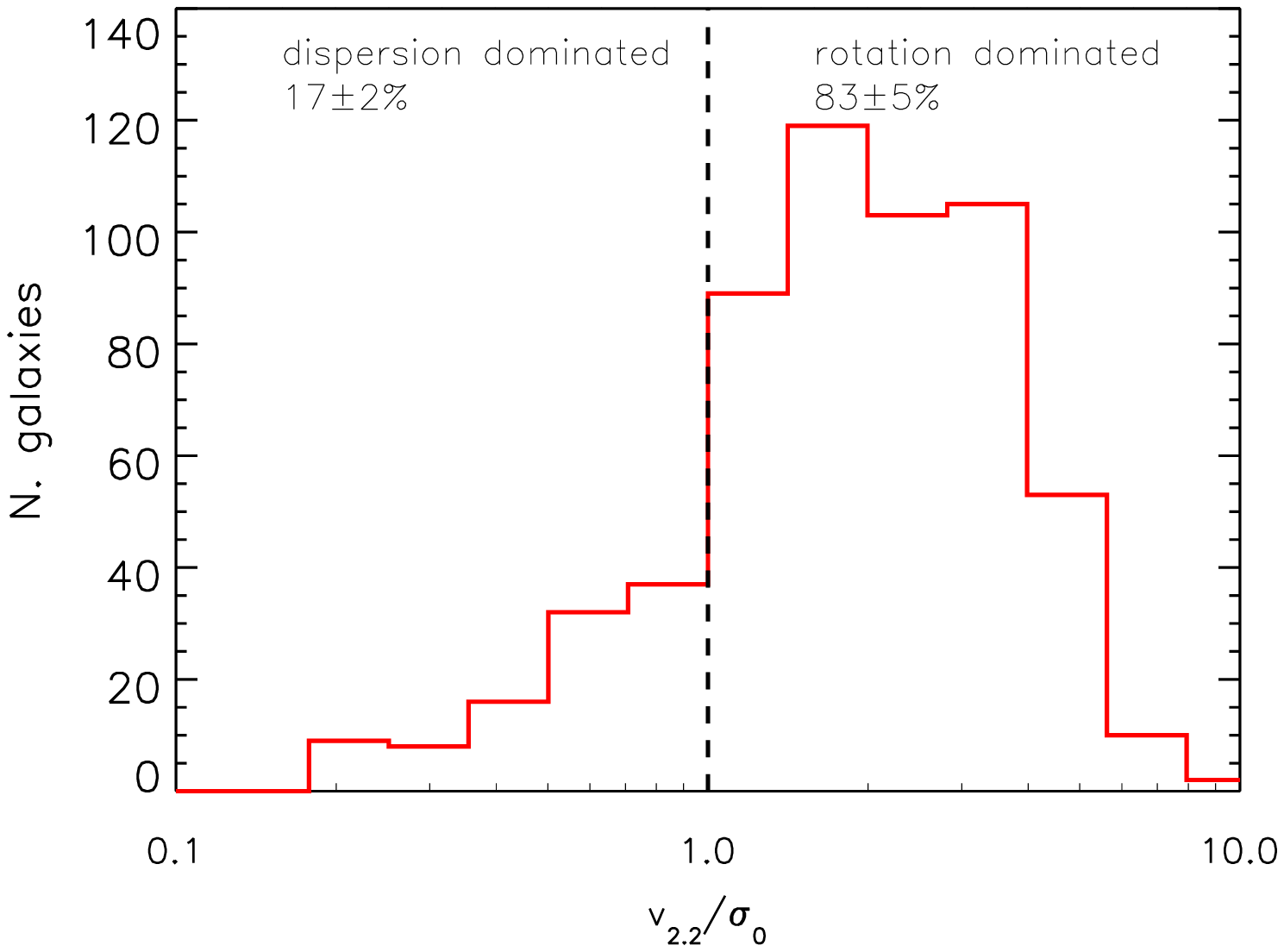}

\caption[]{{\it Left}: The distribution of the intrinsic velocity dispersion, $\sigma_0$, of the galaxies (solid red line). Dashed line is for rotation dominated galaxies only. {\it Centre}: The distribution of the $v_{2.2}$ rotation velocity of the galaxies. {\it Right}: The distribution of the $v_{2.2}/\sigma_0$ of the galaxies. The dashed vertical line at $v_{2.2}/\sigma_0$ represents the delineation we adopt between rotation and dispersion dominated galaxies. This demonstrates that while the KROSS galaxies have a wide range of rotation velocities and dispersions, the majority have dynamics dominated by rotation.}

   \label{fig:sighist}
\end{figure*}

To assess the rotation speed of the galaxies in our sample, we choose to adopt the $v_{2.2}$ parameter, which is the inclination corrected rotation speed at a radius $r_{2.2}$. The radius $r_{2.2}$ is defined as $2.2\times$ the effective (half-light) radius $r_e$. The median $r_{2.2}$ for the KROSS sample is $9.5\pm0.2\rm\, kpc$. The reason for using the velocity at $r_{2.2}$ rather than closer in to the kinematic centre is that we obtain a measure of the rotation velocity as {the rotation curve} flattens \citep{freeman1970} rather than the steep rapidly changing inner part, which will be very sensitive to small uncertainties in radius and therefore difficult to compare between galaxies (see Fig. \ref{fig:vcurve}). {The value of $v_{2.2}$ is an average of the absolute values of the maximum and minimum velocities in the model velocity map at a radius $r_{2.2}$ along the semi-major axis, corrected for inclination}. In Fig. \ref{fig:sighist} we plot the distribution of the $v_{2.2}$ parameter. The median value of $v_{2.2}$ for our sample is $\rm 129 \,km\ s^{-1}$ with a standard deviation (s.d.) of $\rm 88\,km\ s^{-1}$ {and a $21\%$ median uncertainty on individual values of $v_{2.2}$}. We note that for 103 galaxies ($\sim18\%$) the observed rotation curve does not reach $r_{2.2}$ on either side of the kinematic major axis, for these we use the best fitting analytical expression from Eq. \ref{eq:arc} to calculate $v_{2.2}$. This extrapolation is typically only $0.4''$ and so should not affect our results.

The intrinsic velocity dispersion ($\sigma_0$) map of the galaxies is obtained by taking the measured H$\alpha$ linewidth map and {removing the effects of the instrumental resolution and the beam smeared local velocity gradient ($\frac{\Delta V}{ \Delta r}$ of the model velocity field within the PSF radius). This is done using the equation

\begin{equation}
\sigma_{0}^2= \left(\sigma_{obs}-\frac{\Delta V}{ \Delta r}\right)^2 - \sigma_{inst}^2,
\label{eq:beam}
\end{equation}

\noindent where $\sigma_{0}$, \,$\sigma_{obs}$ and $\sigma_{inst}$ are the intrinsic, observed and instrumental values of $\sigma$ respectively. We note that this equation corrects the $\frac{\Delta V}{ \Delta r}$ linearly rather than in quadrature, which we found best recovers the intrinsic velocity dispersion in our simple model (see Appendix A).} To calculate a single intrinsic velocity dispersion for each galaxy we then take the flux weighted average value of the intrinsic velocity dispersion map. In Fig. \ref{fig:sighist} we plot the distribution of the intrinsic velocity dispersion of the galaxies. The median intrinsic velocity dispersion of the whole KROSS sample is $\rm 60\pm43\, (s.d.) \,km\ s^{-1}$, {with a $7\%$ median uncertainty on individual values of $\sigma_{0}$. We note that if we median the $\sigma$ maps rather than flux weight then the median intrinsic velocity dispersion of the KROSS sample is $\rm 53\pm33\, (s.d.) \,km\ s^{-1}$ and this has no effect on our conclusions or any trends with $\sigma_{0}$ that follow}.  See \S\ref{sec:sigdif} on the implications of the technique used to measure $\sigma_0$.

\subsubsection{Rotation and dispersion dominated galaxies}
\label{sec:rotfrac}

By comparing the rotation speed of the galaxies with their intrinsic velocity dispersion we can assess whether their gas dynamics are dominated by ordered rotation or random motions. The dynamical state of the galaxy is assessed using the ratio of $v_{2.2}/\sigma_0$ (\citealt{genzel2006}). The distribution of $v_{2.2}/\sigma_0$ is displayed in Fig. \ref{fig:sighist}. From this we can see that the $v_{2.2}/\sigma_0$ values are typically lower than those of low redshift disc galaxies, which tend to have $v/\sigma\sim5-20$ \citep{epinat2010}. In fact the average value of  $v_{2.2}/\sigma_0$ for the KROSS sample is $2.2\pm1.4 \rm\, (s.d.)$, consistent with the low values measured by \cite{forster2009,swinbank2012,epinat2012} at $z>0.8$. 
 
We choose $v_{2.2}/\sigma_0=1.0$ to make a somewhat crude delineation between galaxies that are supported by ordered rotation and those that are supported by their velocity dispersion. The galaxies with $v_{2.2}/\sigma_0<1.0$ are classified as `dispersion dominated' and those with $v_{2.2}/\sigma_0>1.0$ are `rotation dominated'. We are careful at this stage to use these terms as opposed to `disc' as the KROSS galaxies have significantly lower  $v_{2.2}/\sigma_0$ than local disc galaxies, being more turbulent or dynamically `hotter'. $83\pm5\%$ {of the resolved $\rm H\alpha$ sample} satisfy this rotation dominated criteria. 

The median $\sigma_0$ for the rotation dominated galaxies in KROSS is $\rm 59\pm 32\, (s.d.) \,km\ s^{-1}$, with a standard error (s.e.) of $\rm 2\,km\ s^{-1}$. This is in good agreement with the $z\sim1$ MASSIV SINFONI IFU study \citep{epinat2012} who find an average velocity dispersion of $\rm 62\,km\ s^{-1}$ for their rotators using a similar technique to ours, applying a beam smearing correction and weighted average. Again, we note that if we median the $\sigma$ maps rather than flux weight then the median intrinsic velocity dispersion of the rotation dominated sample is $\rm 52\pm27\, (s.d.) \,km\ s^{-1}$. For a local comparison sample \cite{epinat2010} find low redshift galaxies have velocity dispersions of $\rm 24\pm5\,km\ s^{-1}$, again correcting for the beam smearing of the velocity gradient. This suggests that the velocity dispersion of the gas in rotating star-forming galaxies does indeed increase with redshift. 

\begin{figure*}
   \centering

\includegraphics[scale=0.43, trim=20 0 20 0, clip=true]{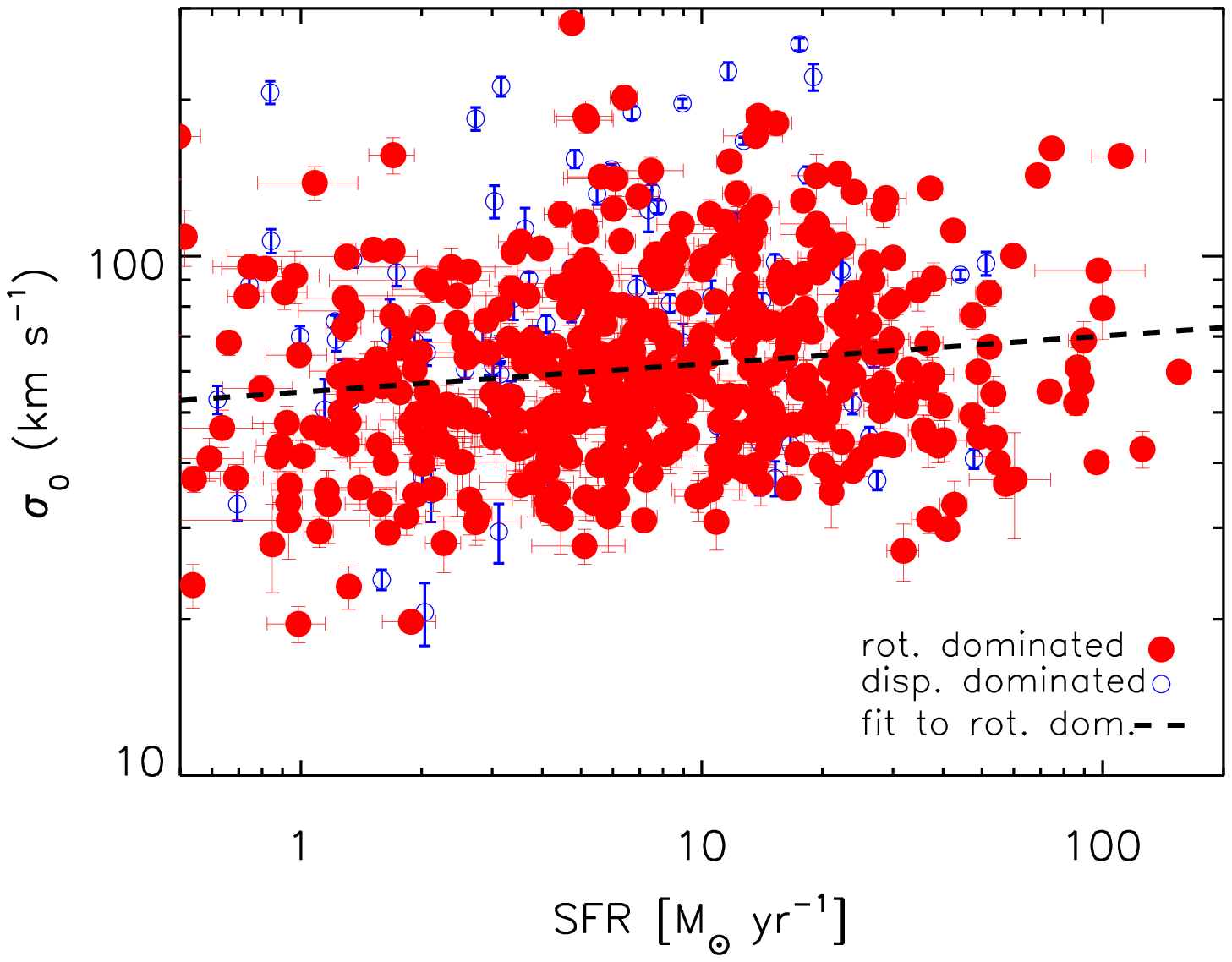}\includegraphics[scale=0.43, trim=85 0 20 0, clip=true]{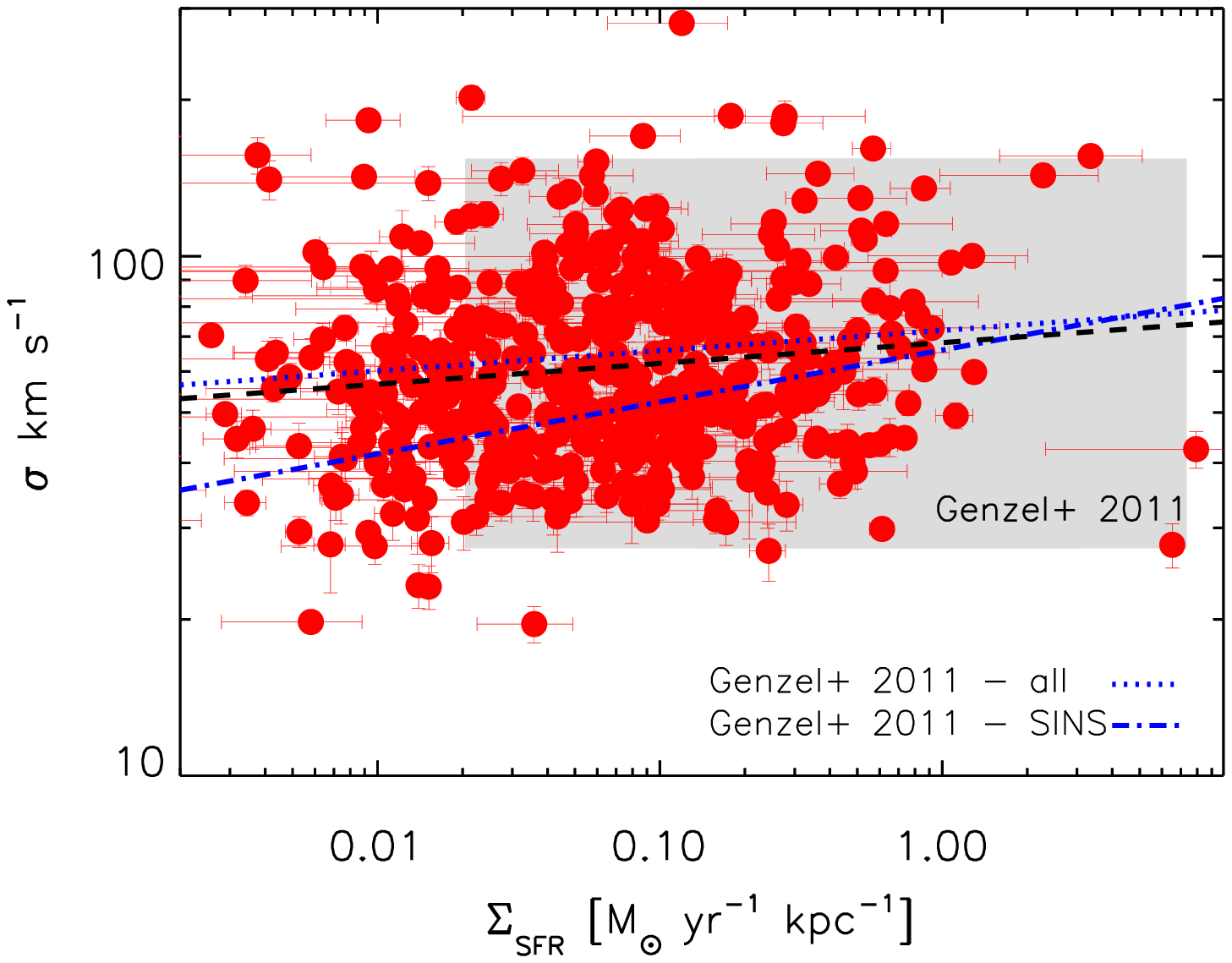}\includegraphics[scale=0.43, trim=85 0 0 0, clip=true]{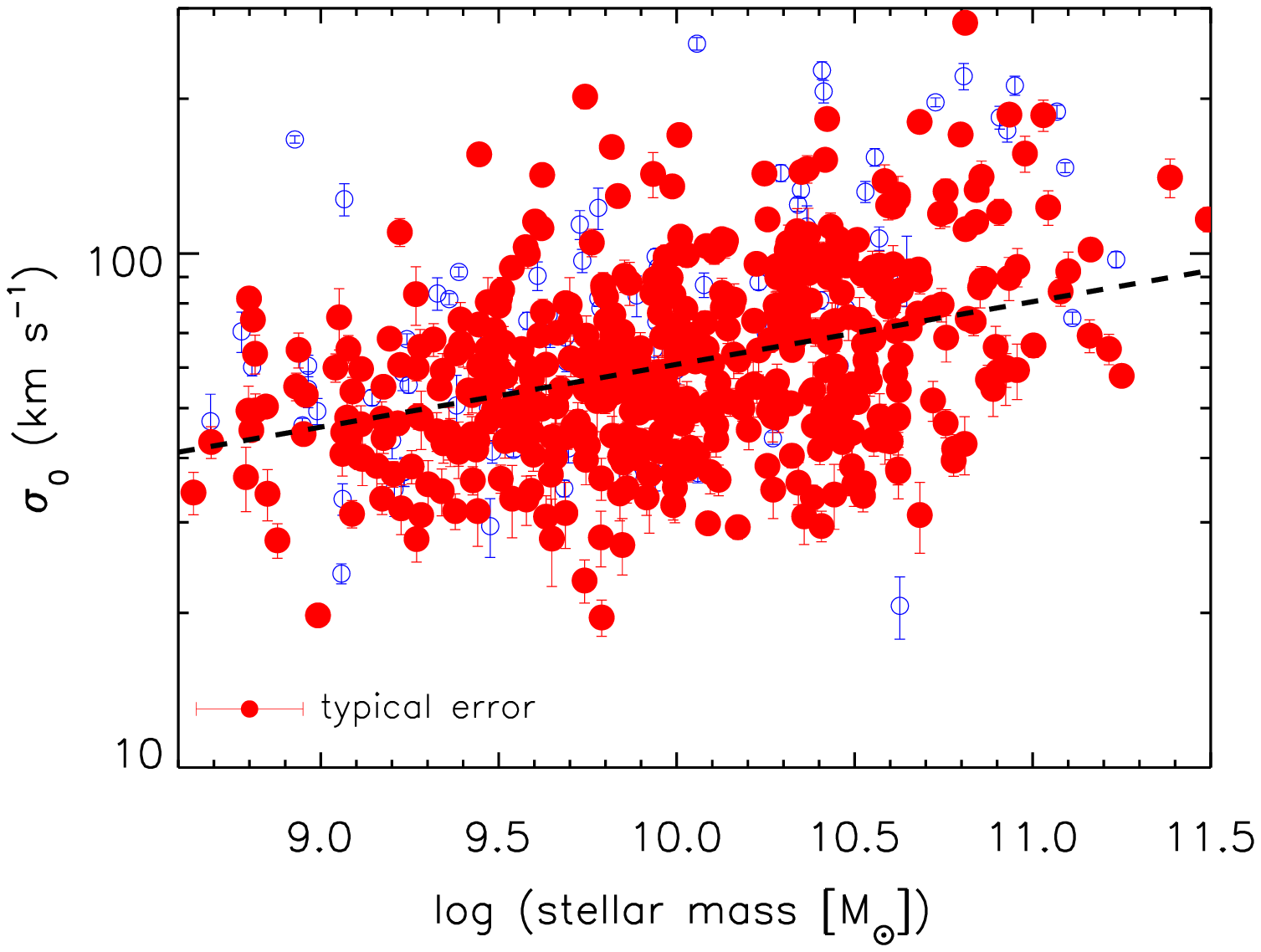}

\caption[]{{\it Left}: We find a weak correlation (black dashed line) between intrinsic velocity dispersion, $\sigma_0$ and SFR for the rotation dominated galaxies (filled red circles). The dispersion dominated galaxies are included for completeness (open blue circles).  {\it Centre}: We find a similarly weak correlation between intrinsic velocity dispersion, $\sigma_0$ and $\rm \Sigma\ _{SFR}$ for the rotation dominated galaxies in agreement with the fits of \cite{genzel2011} (dotted and dot dashed blue lines) whose combined $z>1$ galaxy sample occupies a similar parameter space (grey shaded region). {\it Right}: A moderate correlation is found between $\sigma_0$ and stellar mass perhaps demonstrating that even for rotating galaxies $\sigma_0$ is connected to the potential \citep{kassin2007}. From these plots we conclude that the turbulence in the galaxies is unlikely to be driven by star formation feedback as a stronger correlation would be expected between $\sigma_0$ and SFR or $\rm \Sigma\ _{SFR}$ (e.g. \citealt{lehnert2013,green2014}) rather than stellar mass {(although see \citealt{kimkim2011})}.}

   \label{fig:sigsfr}
\end{figure*}

We note that a correlation has been seen between velocity dispersion and SFR \citep{green2014}, which they attribute to star formation feedback driving the turbulence in the discs, albeit for galaxies with $\rm SFR>10\,M_\odot\,yr^{-1}$. This may also act to confuse any claims of evolution of $\sigma_0$ with redshift due to selection biases to more massive, highly star-forming galaxies. To test whether we also find that SFR feedback is potentially driving the higher turbulence we see in the KROSS sample, in Fig. \ref{fig:sigsfr} we plot SFR against $\sigma_0$ but find only a weak correlation for the rotation dominated galaxies ($\sigma_0\propto \rm SFR^{0.05\pm0.02}$, Pearson's $\rho=0.16$). A moderate correlation is found between $\sigma_0$ and mass for the rotation dominated galaxies ($\sigma_0\propto M_{\star}^{0.12\pm0.01}$, Pearson's $\rho=0.38$). A weak trend between $\sigma_0$ and stellar mass is perhaps not surprising as the gaseous velocity dispersion may be correlated with stellar mass (see \S\ref{sec:mass}) even for galaxies which are rotation dominated \citep{kassin2007}. This in turn correlates with SFR as observed in the main sequence \citep{noeske2007}, which may explain the weak trend with SFR even without any star formation feedback. {There may also be some contribution to this correlation from any uncorrected beam smearing (see Appendix A)}. As the trend between $\sigma_0$ and SFR is very weak and we are studying typical galaxies at $z=1$, {just as those of  \cite{epinat2010} are typical of $z\sim0$}, then there should be little bias in the factor of $\sim2$ evolution in $\sigma_0$ discussed above. 

Rather than the total SFR, the velocity dispersion may correlate with the SFR surface density of the galaxy (e.g. \citealt{genzel2011,lehnert2013}). The SFR surface density, $\rm \Sigma\ _{SFR}$, is a measure of the spatial intensity of the star formation and therefore star formation driven feedback may be stronger in galaxies with a high $\rm \Sigma\ _{SFR}$. The SFR surface density is calculated by dividing half of the total SFR by the area within the $\rm H{\alpha}$ effective radius, $r_e$. In Fig. \ref{fig:sigsfr} we also plot $\sigma_0$ against $\rm \Sigma\ _{SFR}$ and find that $\sigma_0\propto \rm \Sigma\ _{SFR}^{0.04\pm0.01}$. This is in good agreement with the weak relations found by \cite{genzel2011}, whose compilation of SINS and other galaxy samples at $z>1$ occupy a similar parameter space to KROSS ($\sigma_0\propto \rm \Sigma\ _{SFR}^{0.039\pm0.022}$, for all galaxies; $\sigma_0\propto \rm \Sigma\ _{SFR}^{0.1\pm0.04}$ for SINS galaxies and clumps). {\cite{lehnert2013} find a stronger trend, consistent with $\sigma_0\propto \rm \Sigma\ _{SFR}^{0.5}$ as predicted from their simulations. However, they probe a larger dynamic range in $\rm \Sigma\ _{SFR}$ than KROSS, with $\rm \Sigma\ _{SFR}>1 M_{\odot} yr^{-1} kpc^{-2}$ so it is difficult to make a direct comparison.}

The median value of $v_{2.2}$ for the rotation dominated galaxies in KROSS is $\rm 142\pm 85\, (s.d.) \,km\ s^{-1}$. This compares with an average rotation velocity value of $\sim170 \rm \,km\ s^{-1}$ as measured by KMOS$\rm^{3D}$ \citep{kmos3d2014} at $z\sim1-2$. Their value may be higher in part because they are using the average of the maximum and minimum rotation velocities rather than the $v_{2.2}$ value used by KROSS. We note that our sample is also less massive than \cite{kmos3d2014} with our `rotation dominated' galaxies having a median mass $\log(M_{\star} \rm [M_\odot])=10.0$ whereas the average for KMOS$\rm^{3D}$ at $z\sim1$ is 10.65, so lower rotation speeds are expected for KROSS.

\subsubsection{Comparing $v/\sigma$ measurements techniques}
\label{sec:sigdif}

The average velocity dispersion found by KMOS$\rm^{3D}$ \citep{kmos3d2014} for their `disk' galaxies at $z\sim1$ is $\rm 24.9\,km\ s^{-1}$, significantly lower than the value we measure ($\rm 59 \pm 32\, [s.d.]\,km\ s^{-1}$, with a standard error of $\rm 2\,km\ s^{-1}$). This apparent discrepancy between the KROSS and KMOS$\rm^{3D}$ results can be explained by the different methods used to calculate $\sigma_0$. KMOS$\rm^{3D}$ use only data far from the kinematic centre to calculate $\sigma_0$ and do not therefore need to employ a correction for beam smearing as this only affects the regions with high velocity gradients. If we employ a similar method to KMOS$\rm^{3D}$ and extract the median galaxy velocity dispersion beyond $r_{2.2}$, where the rotation curve is flattening, we then find a median $\sigma_{2.2}$ of $\rm 27 \pm 25\, (s.d.) \,km\ s^{-1}$ with a standard error of $\rm 1\,km\ s^{-1}$ (this time defining rotation dominated systems as those with $v_{2.2}/\sigma_{2.2}>1$). This is in much better agreement with the KMOS$\rm^{3D}$ value and confirms that we are observing a similar population. We note that if we correlate $\sigma_{2.2}$ with stellar mass, SFR and  $\rm \Sigma\ _{SFR}$, as was done for $\sigma_{0}$ in \S\ref{sec:rotfrac} and Fig. \ref{fig:sigsfr}, the trends are equally weak. 

If we adopt $v_{2.2}/\sigma_{2.2}$ as our rotation/dispersion dominated criteria instead of $v_{2.2}/\sigma_0$ then the we find an average value of $5.8\pm0.3$, in good agreement with the $\sim5.5$ measured by \cite{kmos3d2014} at $z\sim1$. The percentage of rotation dominated galaxies {in the resolved $\rm H\alpha$ sample} also increases from 83\% to $91\pm6\%$, again in excellent agreement with the 93\% of \cite{kmos3d2014} at $z\sim1$. 

\begin{figure}
   \centering

\includegraphics[scale=0.5, trim=0 0 0 0, clip=true]{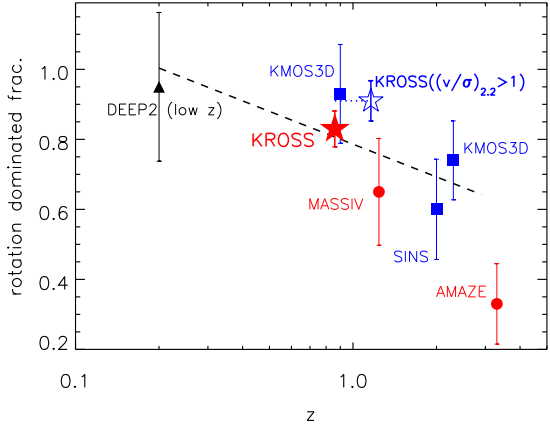}

\caption[]{The rotation dominated fraction plotted against redshift for KROSS and comparison samples. The sample sizes are: KROSS \totres; KMOS$\rm^{3D}$ 90 ($z\sim1$), 101 ($z\sim2$); MASSIV 50; SINS 63; AMAZE 33; and for DEEP2, while the full sample is much larger, we only use the 20 galaxies below $z=0.2$. This plot illustrates the current state of `disc' fraction observations, which suggest a decline with increased redshift of the form ${\rm RDF}\propto z^{-0.2\pm0.1}$ to $z=3$, represented by the dashed line. There are a number of caveats due to {differences in} selection and measurement technique which are discussed in the text. The red circular points are measured in a similar way to the KROSS technique of removing the local beam smeared velocity gradient. The blue squares use measurements of $\sigma$ at large galactocentric radii or are uncorrected for beam smearing. For completeness, we also include the KROSS RDF if $\sigma$ is measured at large galactocentric radius (i.e. $v_{2.2}/\sigma_{2.2}>1$; plotted as a redshift-offset open blue star for clarity and not included in the fit to the decline with $z$). This demonstrates that the KROSS and KMOS$\rm^{3D}$ RDFs agree for the same measurement technique. The black triangle (DEEP2) is from single slit measurements. }
   \label{fig:discfracz}
\end{figure}

In Fig. \ref{fig:discfracz} we plot the RDF (rotation dominated fraction) against redshift for a set of comparable samples. These are DEEP2 \citep{kassin2012}, KMOS$\rm^{3D}$ \citep{kmos3d2014}, MASSIV \citep{epinat2012}, SINS \citep{forster2009} and AMAZE \citep{gnerucci2011}. {Here we define the RDF for MASSIV as the fraction of resolved galaxies with their $v_{max}/\sigma>1$ (65\%), the RDF for SINS as the  fraction of resolved galaxies with their $v_{obs}/(2\sigma_{int})>0.5$ (60\%, from Table 9 of \citealt{forster2009} ) and the RDF for AMAZE as their resolved galaxy `rotator' fraction (33\%), of which all are consistent with their $v_{max}/\sigma_{int}>1$ (see \citealt{gnerucci2011} Fig. 20, although we note that this fraction could be higher if $v_{max}/\sigma_{int}>1$ for any of their `non-rotators')}. Fig. \ref{fig:discfracz} shows that there is a trend between RDF and redshift, such that at higher redshift the number of galaxies with dynamics dominated by ordered rotation reduces. The dependence on redshift for the set of samples below $z=3$ is ${\rm RDF}\propto z^{-0.2\pm0.1}$ and if the AMAZE sample is included this becomes ${\rm RDF}\propto z^{-0.3\pm0.1}$. However, we caution that the rotation dominated galaxies are defined in different ways for different surveys. For example, while KROSS, MASSIV and AMAZE define rotation dominated galaxies in a similar way using a $\sigma_0$ corrected for the local velocity field, KMOS$\rm^{3D}$ use a $\sigma$ at large galactocentric radius, which is typically half the value of the beam smearing correction technique and thus a $v/\sigma>1$ selection increases the number of rotation dominated systems (as discussed above). For completeness, we also include the KROSS RDF if  the $v_{2.2}/\sigma_{2.2}>1$ definition is used. DEEP2 is based on slit spectroscopy so again may not be directly comparable. 

We caution that these samples are not selected in the same way with potential biases to more massive or more star-forming systems at $z>1$ {(see Fig. \ref{fig:sfrhist})}. If, for example, these are galaxies selected from a sample biased above the main sequence (high specific SFR, [$\rm sSFR=SFR/M_{\star}$]) then they are significantly more likely to be merging or interacting systems \citep{stott2013} and thus the fraction of ordered rotators will be lower. Fig. \ref{fig:discfracz} can therefore only be considered illustrative of the current state of disc/rotation dominated fraction evolution.

\subsection{Dynamical mass}
\label{sec:mass}

The resolved kinematics of the galaxies can be used to calculate their dynamical mass within a given radius. The majority of the galaxies are rotation dominated ($v_{2.2}/\sigma_0>1$) and as such we can calculate the dynamical mass, assuming a spheroidal distribution, via the Keplerian formula: 

\begin{equation}
M_{dyn}(r<r_{2.2})=\frac{v_{2.2}^2 r_{2.2}}{G},
\end{equation}

\noindent where $r_{2.2}$ and $v_{2.2}$ are defined as in \S\ref{sec:dyn} and $G$ is the gravitational constant. We choose to define the dynamical mass within $r_{2.2}$ as this radius typically falls well into the flattened outer portion of rotation curve in disc galaxies as it is significantly larger (by a factor of $\sim2-4$) than the kinematic scale radius ($r_{t}$ from Eq. \ref{eq:arc}), where the velocity is rapidly changing.

For the galaxies that are velocity dispersion dominated ($v_{2.2}/\sigma_0<1$) we instead calculate the dynamical mass from the virial theorem, via the formula:

\begin{equation}
\label{eq:sig}
M_{dyn}(r<r_{2.2})=\frac{\alpha\ \sigma_0^2 r_{2.2}}{G},
\end{equation}

\noindent where $\sigma_0$ is the intrinsic velocity dispersion and $\alpha$ is a constant which typically has a range of values from $2.5-8.5$ depending on geometry, whether the total mass or the mass within $r$ is required and the definition of $r$ (see \citealt{agnello2014} and references therein). We choose to estimate the value of $\alpha$ appropriate for $r_{2.2}$ by using the value that gives the same median dynamical mass to stellar mass ratio as for the rotation dominated galaxies. This gives $\alpha=3.4$ which we use in Eq. \ref{eq:sig} to estimate the dynamical masses of our dispersion dominated galaxies. From these calculations the median dynamical mass of the whole KROSS sample is found to be $\log(\rm M_{dyn} [M_\odot])=10.6\pm0.6$, {with a $39\%$ median uncertainty on individual values}.

\subsection{Gas and dark matter fractions}
\label{sec:fgasdm}

\begin{figure}
   \centering

\includegraphics[scale=0.5, trim=0 0 0 0, clip=true]{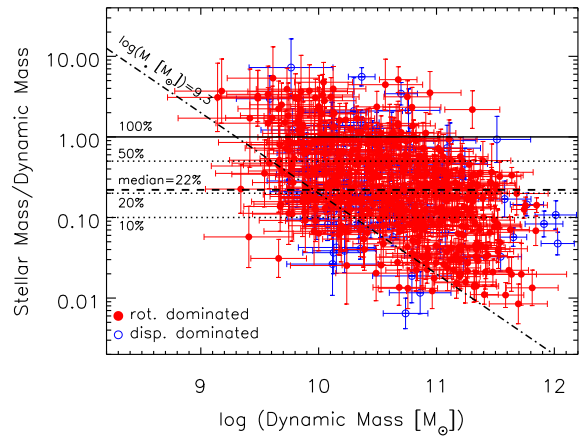}

  \caption[]{The stellar to dynamical mass ratio of the galaxies plotted against dynamical mass. The red filled points are the rotation dominated galaxies and the blue open circles represent those that are dispersion dominated. The solid horizontal line is a one to one ratio and the dotted lines are the 50, 20 and 10\% mass lines. The dashed line is the median ratio of stellar mass to dynamical mass of 22\%. This means that the KROSS galaxies are composed of $78\%$ non-stellar material on average. A caveat is the diagonal dot-dashed line which represents the approximate stellar mass limit of our selection criteria discussed in \S\ref{sec:dat}, $\log(M_{\star} \rm [M_\odot])=9.3$. This selection effect could potentially bias the average stellar mass to dynamical mass ratio to higher values. }
  
   \label{fig:massmass}
\end{figure}

Now that we have calculated the dynamical mass we plot this quantity against the stellar to dynamical mass ratio of the galaxies (Fig. \ref{fig:massmass}). The stellar mass is measured within a $2''$ diameter aperture and the median $r_{2.2}$ of KROSS is $9.5\pm0.2\,\rm kpc$, which corresponds to a diameter of $2.4''$ at the median redshift of the survey, $z=0.85$. To assess whether an aperture correction should be applied we perform the same test used in \S\ref{sec:msfr} and find that for an exponential profile galaxy with the KROSS average half-light radius ($0.6''$), $2.0''$ should contain 95\% of the light, in typical ground-based seeing ($0.7''$). So assuming mass follows light then we choose not to apply any aperture corrections. {The justification for using the half-light radius of the $\rm H{\alpha}$ as a proxy for that of the stellar light is that the $\rm H{\alpha}$ extent is found to be systematically larger than that of the stellar light for star-forming galaxies at these redshifts \citep{nelson2012}.}
The median ratio of the stellar mass of the galaxies to that of the dynamical mass for the entire KROSS sample is $22\pm11\%$, {although we note that this is driven by the rotation dominated galaxies only as the dispersion dominated galaxies have a fixed median stellar to dynamical mass ratio (see \S\ref{sec:mass})}. However, the approximate KROSS stellar mass limit of $\log(M_{\star} \rm [M_\odot])\sim9.3$ discussed in \S\ref{sec:dat} (represented by the diagonal line on Fig. \ref{fig:massmass}) means that for galaxies with dynamical masses of $\log(\rm M_{dyn} [M_\odot])\lesssim10.5$ the full range of possible stellar masses may not be sampled. This could act to bias the average stellar mass to dynamical mass ratio to a higher value.

{There are 80 galaxies with larger stellar mass than dynamical mass, representing $14\%$ of the total sample. We note that this fraction reduces to $8\%$ ($4\%$) if we assume the ratio of their stellar to dynamical mass is $1(2) \sigma$ lower so this may be mainly due to measurement uncertainty.} The dynamical mass represents the total mass of the system within 2.2\ $r_e$ meaning that, on average, $78\pm11\%$ of the mass within this radius is not composed of stellar material. This non-stellar material will be composed of gas and, at this radius, a large contribution from dark matter. To illustrate this we estimate the dark matter fraction $f_{DM}$ within $r_{2.2}$ by studying the output of the state-of-the-art Evolution and Assembly of GaLaxies and their Environments (EAGLE) hydrodynamic simulation \citep{schaller2014,schaye2014,crain2015,mcalpine2015}. For {109} EAGLE galaxies at $z=0.9$ with $\log(M_{\star} \rm [M_\odot])=9.7-10.3$ and $\rm 5 < SFR < 10 \,M_\odot yr^{-1}$ (to approximately match KROSS) the $f_{DM, E}$ within $10\,\rm kpc$ {(best available aperture for comparison with the average $r_{2.2}$ of KROSS)} ranges from $33-77\%$ with a median value of $67\pm8\%$. Assuming this average dark matter fraction from EAGLE then we can approximate an average gas mass to total mass fraction for the KROSS sample of $f_{g, dyn, t}\sim11\%$ (giving an average gas mass to baryonic mass fraction of $f_{g, dyn, b}\sim33\%$).

To obtain an alternative estimate of the gas fraction, using an orthogonal method based on KROSS observables, we can invert the Kennicutt Schmidt relation (KSR, \citealt{kennicutt1998}). This is done by using the formula:

\begin{equation}
\label{eq:ks}
\Sigma_{\rm gas}=\left(\frac{\Sigma_{\rm SFR}}{2.5 \times10^{-4}}\right)^{0.714},
\end{equation}

\noindent where $\Sigma_{\rm SFR}$ is the SFR surface density ($\rm M_\odot yr^{-1} kpc^{-2}$), which we calculate by dividing half of the total SFR by the area within the $\rm H{\alpha}$ effective radius, $r_e$. The quantity $\Sigma_{\rm gas}$ is the gas surface density ($\rm M_\odot pc^{-2}$) which can be converted to a total gas mass by multiplying by the area and then by a factor of two (as we only considered the SFR density within $r_e$). For KROSS the median gas surface density is found to be $54\pm5\rm \,M_\odot pc^{-2}$. The average gas to total mass fraction, $f_{g, KS, t}$ is found to be $13\pm5\%$, in agreement with the crude estimate $f_{g, dyn, t}\sim11\%$ assuming the EAGLE dark matter fraction above.

However, the gas fraction is more commonly expressed as a ratio of the gas mass to the baryonic mass, in which case 

\begin{equation}
f_{g, KS, b}=\frac{M_{g,KS}}{M_{\star}+M_{g,KS}}, 
\label{eq:ksr}
\end{equation}

\noindent where $M_{g,KS}$ is the gas mass inferred from inverting the KSR, the median $f_{g, KS, b}=35\pm7\%$. The median gas to baryonic mass fraction within $10\,\rm kpc$ extracted from the EAGLE simulation galaxies, $f_{g, E, b}=40\pm9\%$, is in good agreement with the median value inferred from our measurements but we note that EAGLE has the KSR built into it so this may be unsurprising for a galaxy sample of similar SFR. {The gas fraction becomes higher for progressively lower mass galaxies with a median $f_{g, KS, b}= 0.17\pm0.03$ for galaxies with $\log(M_{\star} \rm [ M_\odot])>10$ and a median $f_{g, KS, b}= 0.64\pm0.13$ for $\log(M_{\star} \rm  [M_\odot])<10$. This is unsurprising as for a sample spanning a range of masses with an approximate star formation limit of SFR$\, \sim \rm 1\,M_\odot yr^{-1}$ (corresponding to our H$\alpha$ flux detection threshold) then the lower mass galaxies will be biased to those with higher sSFR and hence higher gas fractions.}

At $z\gtrsim1$ the large column densities and interstellar pressures in star-forming galaxies (with $\Sigma_{\rm gas}\gg10\rm M_\odot pc^{-2}$, for KROSS  $\Sigma_{\rm gas}=54\pm5\rm \,M_\odot pc^{-2}$) mean that most of the cold interstellar gas is likely in molecular form and the contribution of atomic hydrogen can be neglected \citep{blitz2006,tacconi2010}. In which case the gas fractions calculated by either the dynamical or KSR method are in good agreement with those found through the study of molecular gas fractions at this redshift (\citealt{daddi2010,tacconi2010,swinbank2012clump}, see \S\ref{sec:disc}).

By assuming that all of the gas is available as fuel for star formation and the current SFR remains constant we can calculate a gas depletion timescale ($t_{\rm dep}$). This is simply the gas mass divided by the star formation rate

\begin{equation}
\label{eq:tdep}
t_{\rm dep}=\frac{ M_{ g, KS}}{\rm SFR}.
\end{equation}
  
\noindent We find an average $t_{\rm dep}$ of $10^{8.9\pm0.2}\rm \,yr$, meaning that, assuming no additional gas enters the system, then these galaxies will use up their fuel for star formation within $0.8\pm0.4\rm\, Gyr$. This is in full agreement with the average found by \cite{saintonge2011b} in nearby galaxies of $\sim 1\rm\, Gyr$.

We note that combining the gas mass from inverting the KSR with the stellar mass from SED fitting and the total dynamical mass gives a measure of the dark matter content, as  
\begin{equation}
M_{dyn}=M_{\star}+M_{gas}+M_{DM}.
\end{equation}

\noindent Using this formula, we find that within $r_{2.2}$ the average dark matter fraction $f_{DM}=65\pm12\%$ for the KROSS sample galaxies. {The three mass components were not all calculated in the exact same aperture, which may be responsible for some of the scatter in $f_{DM}$} . This is in good agreement with the $z=0.9$ predictions of the EAGLE simulation within $\rm10\,kpc$, $f_{DM, E}=67\pm8\%$ as discussed above. This $f_{DM}$ is also in good agreement with that derived from local disc galaxy observations, again within $r_{2.2}$ ($f_{DM}=68\%,\, $\citealt{courteau2015})

\subsection{Disc stability}
\label{sec:stab}

We have shown that typical rotation dominated $z\sim1$ star-forming galaxies have high gas fractions and larger values of $\sigma_0$ compared with galaxies in the local Universe. It is also possible to assess the stability of these rotating, turbulent discs through the Toomre $Q$ parameter \citep{toomre1964}. 

We calculate Q for our rotation dominated galaxies ($v_{2.2}/\sigma_0>1$) by using the formula

\begin{equation}
\label{eq:q}
Q=\frac{\sigma_0}{v_{2.2}}\ \frac{a}{f_{gas}},
\end{equation}

\noindent where $a$ is a constant, which for the flattened region of the rotation curve at $\sim r_{2.2}$ has a value $a=\sqrt2$ \citep{genzel2011,kmos3d2014}. The quantity $f_{gas}=f_{g, KS, b}$ is the gas fraction we calculate through inverting the KSR in Eq. \ref{eq:ksr}. If the value of $Q$ is high ($Q>1$) then the disc is stable, if it is low ($Q<1$) then the disc is unstable to gravitational fragmentation but at $Q=1$ the disc is thought to be marginally stable i.e. on the verge of becoming unstable. 

\begin{figure}
   \centering

\includegraphics[scale=0.5, trim=0 0 0 0, clip=true]{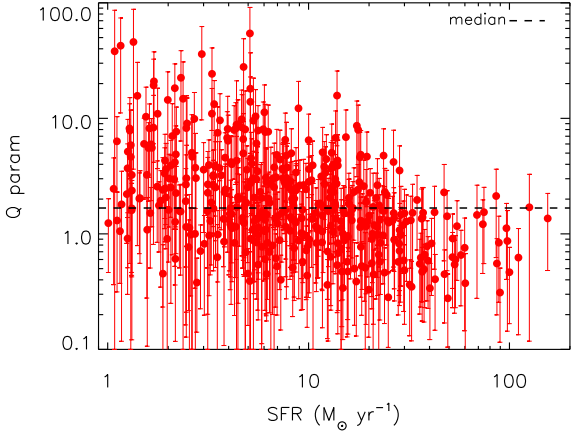}

\caption[]{The Toomre Q parameter of the KROSS rotation dominated galaxies plotted against their SFR. The majority of the galaxies are consistent with being marginally stable/unstable discs with $Q=1$. The median value is $Q=1.7\pm0.4$. Such gravitational instabilities are thought to induce galaxy-wide star formation in disc galaxies \citep{elmegreen2002,li2005} and are therefore likely to be responsible for the increased SFRD at $z=1$. }
   \label{fig:qsfr}
\end{figure}

In Fig. \ref{fig:qsfr} we plot the $Q$ parameter for our galaxies against the SFR and see no trend with this or any other observable. We find that the majority of the individual galaxies can be considered to have gaseous discs consistent with $Q\leq1$ within their error bars ($75[99]\%$ within $1[2]\sigma$) and the median value of $Q$ is found to be $1.7\pm0.4$. This indicates that the gaseous discs of these galaxies can be considered to be marginally (un)stable, as has been seen before for smaller samples at similar redshifts \citep{forster2006,genzel2011,swinbank2012clump}. It is such gravitational instabilities that are thought to induce the star formation in disc galaxies by causing the gas to clump and condense \citep{elmegreen2002,li2005}. Other authors such as \cite{kmos3d2014} have inverted Eq. \ref{eq:q} by assuming that $Q=1$ and used it to calculate the gas fraction for the galaxies. Our results demonstrate that this is a valid approach.

\section{Discussion}
\label{sec:disc}

The main driver for KROSS is to survey main sequence star-forming galaxies at $z=1$ to understand typical galaxies at the critical period towards the end of the peak in the universal SFRD \citep{lilly1996,madau1996,hb2006,sobral2013}. This is needed because previous studies, without the benefit of the multiplexing ability of KMOS, had instead concentrated on relatively small samples of potentially highly selected galaxies and thus their conclusions may not be applicable to the general population. 

With KROSS we can now make some general comments on the nature of typical star-forming galaxies at $z=1$. Defining galaxies with $v_{2.2}/\sigma_0>1$ as rotation dominated we find that they account for 83\% of the KROSS galaxies. By comparison with low redshift samples we find that the average intrinsic velocity dispersion, $\sigma_0$, of these rotating galaxies is larger by a factor $\sim2$ at $z\sim1$. The $v_{2.2}/\sigma_0$ values for KROSS (regardless of the definition of $\sigma_0$) are therefore significantly lower on average than their low redshift counterparts meaning that typical disc-like galaxies at this epoch really are dynamically hotter and can be thought of as turbulent (see also \citealt{forster2006,swinbank2012}). {Unlike \cite{lehnert2013} and \cite{green2014} we find no strong correlation between $\sigma_0$ and SFR or $\Sigma_{\rm SFR}$ compared to the trend with stellar mass, suggesting that star formation feedback is not the main driver of the turbulence in typical $z=1$ galaxies (see also \citealt{genzel2011})}. {However, our sample does not probe the high $\Sigma_{\rm SFR}$ values of \cite{lehnert2013}. We also note that some star formation feedback models do not expect an observed correlation \citep{kimkim2011}} 

The median KROSS {gas to baryonic mass} fraction, from inverting the KSR, at $z=1$ is $f_{g, KS, b}=35\pm7\%$ so we now compare this to literature measurements at a similar epoch. As discussed in \S\ref{sec:fgasdm}, most of the cold interstellar gas is likely in molecular form and the contribution of atomic hydrogen can be neglected \citep{blitz2006,tacconi2010}. If this is the case then we can directly compare our gas masses to those presented in \cite{tacconi2010} who use CO measurements to infer molecular gas fractions to total baryonic mass of $34\pm4\%$ for galaxies with $\log(M_{\star} \rm [M_\odot])>10$ at $z\sim1.2$, which is clearly in good agreement. We note that if we reduce our sample to the most massive and highest SFR systems ($\log[M_{\star} \rm (M_\odot)]>10$ and $\rm SFR>20\,M_\odot\,yr^{-1}$), more comparable to the \cite{tacconi2010} selection, then our value is still consistent with $f_{g, KS, b}=38\pm8\%$.

The KROSS results demonstrate that galaxies at this epoch appear to have high {gas to baryonic mass} fractions ($35\%$) comparable to those measured through CO observations at $z\sim1$ but how do these results compare to galaxies at $z=0$? 
If we use the gas masses from \cite{leroy2008}, we obtain a {median} gas {to baryonic mass} fraction of $22\pm4\%$ for {19} local galaxies in the same stellar mass range as KROSS {($8.7<\log[M_{\star}]<10.9$). Although we note that they measure their stellar and gas masses within a median radius of $15.6\pm1.1\rm kpc$, which is on average a factor of 1.6 larger than the median KROSS $r_{2.2}$}. By taking the KROSS gas fractions inferred by inverting the KSR ($f_{g, KS, b}=35\pm7\%$) and comparing to the \cite{leroy2008} then there is little evidence that a strong decrease (only $1.6\sigma$) in the total gas content of star-forming galaxies has occurred since $z=1$. However, from the data presented in \cite{leroy2008} and \cite{saintonge2011} we find average molecular gas fractions of $4\pm1\%$ and $9\pm5\%$ for local galaxies respectively, so as discussed above, if we assume the KROSS gas fraction is dominated by molecular gas then this has decreased by a factor of $\sim4-9$ since $z=1$. We note that in a forthcoming paper we will present ALMA observations to infer the molecular gas mass for a subset of the KROSS galaxies using an orthogonal method.

In an idealised scenario, assuming no further gas is added, then at constant star formation the KROSS galaxies will use up their fuel in $t_{\rm dep}\sim0.8\pm0.4\,\rm Gyr$ (i.e. by $z=0.7$) and would therefore now be passive systems. However, this is probably unrealistic as further gas accretion is likely to occur and the star formation may be episodic. This depletion timescale is in agreement with the average found by \cite{saintonge2011b} in nearby galaxies of $\sim 1\rm\, Gyr$. 

By combining the gas mass from inverting the KSR with the stellar mass and dynamical mass we can estimate the dark matter fraction of the KROSS galaxies within $2.2\,r_e$ ($\sim9.5\rm\,kpc$). A median dark matter fraction of $f_{DM}=65\pm12\%$ is inferred which is in excellent agreement with the median value extracted at the same redshift from the EAGLE hydrodynamical simulation of $f_{DM, E}=67\pm8\%$ and with that of local disc galaxies ($f_{DM}=68\%$, within $2.2\,r_e$, \,\citealt{courteau2015}).

The stability of the discs is measured using the Toomre $Q$ parameter, with the result that the rotation dominated, typical star-forming galaxies at $z\sim1$ are consistent with being marginally (un)stable ($Q\sim1$, see also \citealt{forster2006,genzel2011,kmos3d2014,swinbank2012clump}). Instabilities such as these are thought to create high gas density regions, inducing galaxy-wide star formation \citep{elmegreen2002,li2005} and are therefore likely to be the reason for the increase in the average sSFR of the main sequence to $z=1$ \citep{elbaz2011} and the elevated SFRD of the Universe. 

Taking all of this evidence together we can state that the gaseous discs of normal main sequence star-forming galaxies at $z=1$ are significantly different to those in the local Universe. They are dynamically much hotter, on the verge of fragmenting and are likely dominated by molecular gas perhaps fuelled by efficient cold accretion at this epoch \citep{keres2005,dekel2009}. While the accretion can generate the initial turbulence, \cite{elmegreen2010} find that to sustain the turbulence disc instabilities and star formation driven feedback are required. There is clear evidence of instabilities from the $Q\sim1$ measurement but the lack of a strong correlation between $\sigma_0$ and SFR or $\Sigma_{\rm SFR}$ that we find suggests that feedback may not be a dominant contributor to their turbulence. We speculate that the discs may be kept turbulent through ongoing disc instabilities or continuous accretion of cold and clumpy gas from the cosmic web \citep{keres2005,dekel2009}.

\section{Summary}
\label{sec:sum}

With \totres\,spatially resolved galaxies, KROSS constitutes the largest near-infrared IFU survey of $z\sim1$ galaxies. We have demonstrated the KROSS selection technique to be very successful in that we detect $\rm H\alpha$ emission in $90\%$ of the galaxies we observe, of which 81\% are spatially resolved.\\ 
The key results from this paper are as follows:

\begin{enumerate}

\item At $z\sim1$ the majority of star-forming galaxies are rotationally supported although they are dynamically hotter than their local counterparts with on average higher velocity dispersions and therefore significantly lower values of $v/\sigma$ (average KROSS $v_{2.2}/\sigma_0=2.2\pm1.4$).

\item Typical star-forming galaxies at $z\sim1$ are gas rich, with gas inferred to account for $\sim35\%$ of the baryons on average.

\item The rotation dominated galaxies are all consistent with being marginally (un)stable as indicated by their consistency with a Toomre parameter $Q\sim1$.

\item The intrinsic velocity dispersion is not strongly correlated with star formation rate or star formation rate surface density, which may indicate that star formation feedback is not the main driver of the turbulence in typical star-forming galaxies.

\item When comparing KROSS with other samples from the literature the fraction of rotation-dominated galaxies appears to decrease with redshift, although this is subject to selection effects and disparities in measurement technique.

\item Within $2.2\,r_{e}$ star-forming galaxies at $z\sim1$ have rotation dominated by dark matter with an average fraction $f_{DM}=65\pm12\%$, in good agreement with EAGLE hydrodynamic simulation.

\end{enumerate}

From these results we conclude that the elevated SFR of typical star-forming galaxies and SFRD of Universe found at $z\gtrsim1$ must be in-part driven by the high (probably molecular, \citealt{tacconi2010}) gas fractions and the gravitational instabilities within their $Q\sim1$ discs. This means that there is sufficient fuel and a mechanism for it fragment and condense into star-forming regions in order to sustain the enhanced SFRs seen at this epoch.

\section*{Acknowledgments}

{We thank the anonymous referee for their detailed comments, which have improved the clarity of this paper.} JPS, AMS, RGB, CMH and IRS acknowledge support from STFC (ST/I001573/1 and ST/L00075X/1). JPS also gratefully acknowledges support from a Hintze Research Fellowship. IRS acknowledges support from an ERC Advanced Investigator programme DUSTYGAL 321334 and a Royal Society/Wolfson Merit Award. AJB gratefully acknowledges the hospitality of the Research School of Astronomy \& Astrophysics at the  Australian National University, Mount Stromlo, Canberra where some of this work was done under the Distinguished Visitor scheme. DS acknowledges financial support from the Netherlands Organisation for Scientific research (NWO) through a Veni fellowship and from FCT through a FCT Investigator Starting Grant and Start-up Grant (IF/01154/2012/CP0189/CT0010). PNB is grateful for support from STFC
 (ST/M001229/1)

We thank Holly Elbert, Timothy Green and Laura Prichard for their observations and Alex Karim for providing main sequence fits. We also thank Matthieu Schaller for discussions on the EAGLE simulation quantities and Omar Almaini for discussions on the UDS. 

Based on observations made with ESO Telescopes at the La Silla Paranal Observatory under the programme IDs 60.A-9460, 092.B-0538, 093.B-0106, 094.B-0061 and 095.B-0035. This research uses data from the VIMOS VLT Deep Survey, obtained from the VVDS database operated by Cesam, Laboratoire d'Astrophysique de Marseille, France. This paper uses data from the VIMOS Public Extragalactic Redshift Survey (VIPERS). VIPERS has been performed using the ESO Very Large Telescope, under the ``Large Programme" 182.A-0886. The participating institutions and funding agencies are listed at http://vipers.inaf.it. This paper uses data from zCOSMOS which is based on observations made with ESO Telescopes at the La Silla or Paranal Observatories under programme ID 175.A-0839. We acknowledge the Virgo Consortium for making their simulation data available. The EAGLE simulations were performed using the DiRAC-2 facility at Durham, managed by the ICC, and the PRACE facility Curie based in France at TGCC, CEA, Bruyres-le-Ch‰tel. {This work is based in part on data obtained as part of the UKIRT Infrared Deep Sky Survey.}

\bibliographystyle{mn2e}
\bibliography{kross}

\appendix
\section{Beam smearing}
\label{sec:beam}

{Beam smearing describes the contribution to the local velocity dispersion from the local velocity gradient smeared out by the PSF. This is a significant contributor to the individual spaxel velocity dispersion measurements for IFU studies of distant galaxies in natural seeing. Here we describe our method to remove beam smearing from the KROSS galaxy velocity dispersion maps. 

To assess techniques to remove beam smearing we set up a simple simulation, ignoring instrumental $\sigma$, as follows:

\begin{enumerate}

\item An artificial datacube is created with $30\times30$ spaxels (to approximate a KROSS KMOS datacube oversampled to $0.1''$ pixels), each containing a Gaussian emission line, all with same intrinsic $\sigma_{int}=60\rm km\,s^{-1}$ (to approximate the KROSS average).

\item These emission lines are offset from each other in velocity/wavelength by applying a constant velocity gradient $\left(\frac{\Delta V}{\Delta R}\right)$ in the spatial direction of the $x$-axis.

\item A circularly symmetric S\'{e}rsic $n=1$ light profile with $r_{e}=6$\,pixels is centred in the data cube such that the emission line fluxes diminish with radius from the centre, to model an ideal disc galaxy with the same median size as the KROSS galaxies.

\item This S\'{e}rsic profile is convolved with a Gaussian PSF of 7 pixels to replicate the average KROSS seeing of $0.7''$.

\item The light profile of a given pixel is taken as a sum of its own intrinsic line profile with a relative flux value of 1 plus the intrinsic line profiles and offsets in velocity of all of the other spaxels in the datacube scaled to their fluxes from the S\'{e}rsic profile convolved with the PSF. This profile is fitted with a Gaussian with a velocity dispersion $\sigma_{obs}$ (see Fig. \ref{fig:beam}).

\end{enumerate}

We then attempt to remove the effect of this beam smearing and recover the input $\sigma_{int}$ by either removing the $\Delta V/\Delta R$ in a linear subtraction: 

\begin{equation}
\label{eq:lin}
\sigma_{rec}=\sigma_{obs} - \frac{\Delta V}{\Delta R},
\end{equation}

\noindent where $\sigma_{rec}$ is the recovered $\sigma$, or in quadrature:

\begin{equation}
\label{eq:quad}
\sigma_{rec}^2=\sigma_{obs}^2 - \left(\frac{\Delta V}{\Delta R}\right)^2.
\end{equation}

The result of these tests for the central pixel can be seen in Fig. \ref{fig:beam}. The linear removal of the beam smearing is clearly an improvement over the removal in quadrature. We now obtain the maximum spaxel value of $\Delta V/\Delta R$ for each of the KROSS velocity maps. The median maximum $\Delta V/\Delta R=13.4 \rm \,km\,s^{-1} spaxel^{-1}$, which we use to estimate the typical ratio of the recovered to intrinsic velocity dispersion, ($\sigma_{rec}/\sigma_{int}$). This results in only a $20\%$ residual for the linear removal compared with a $40\%$ excess when removed in quadrature. We note that the beam smearing correction is improved at increasing galacto-centric radius as the contribution from the bight central pixels diminishes. To quantify this, at 7 pixels (1 PSF FWHM) from the centre of our model galaxy, the linear and quadrature residuals improve to $12\%$ and $33\%$ respectively. 

Further evidence that this correction is working come from performing the same test as \cite{epinat2010}, looking for a correlation between $\sigma$ and $v_{2.2}\sin i$. For the KROSS data without the correction a linear fit gives $\sigma_{obs}\propto (v_{2.2}\sin i)^{0.19}$ with Pearson's $r=0.26$. With the beam smearing correction applied $\sigma_{obs}\propto (v_{2.2}\sin i)^{0.11}$ and $r=0.17$. The linear beam smearing correction is simple and, we believe, a reasonable technique to use with high-$z$ IFU observations in natural seeing and so we adopt it for this paper. The full equation incorporating the instrumental $\sigma$ can be found in Eq. \ref{eq:beam}.

\begin{figure}
   \centering

\includegraphics[scale=0.5, trim=0 0 0 0, clip=true]{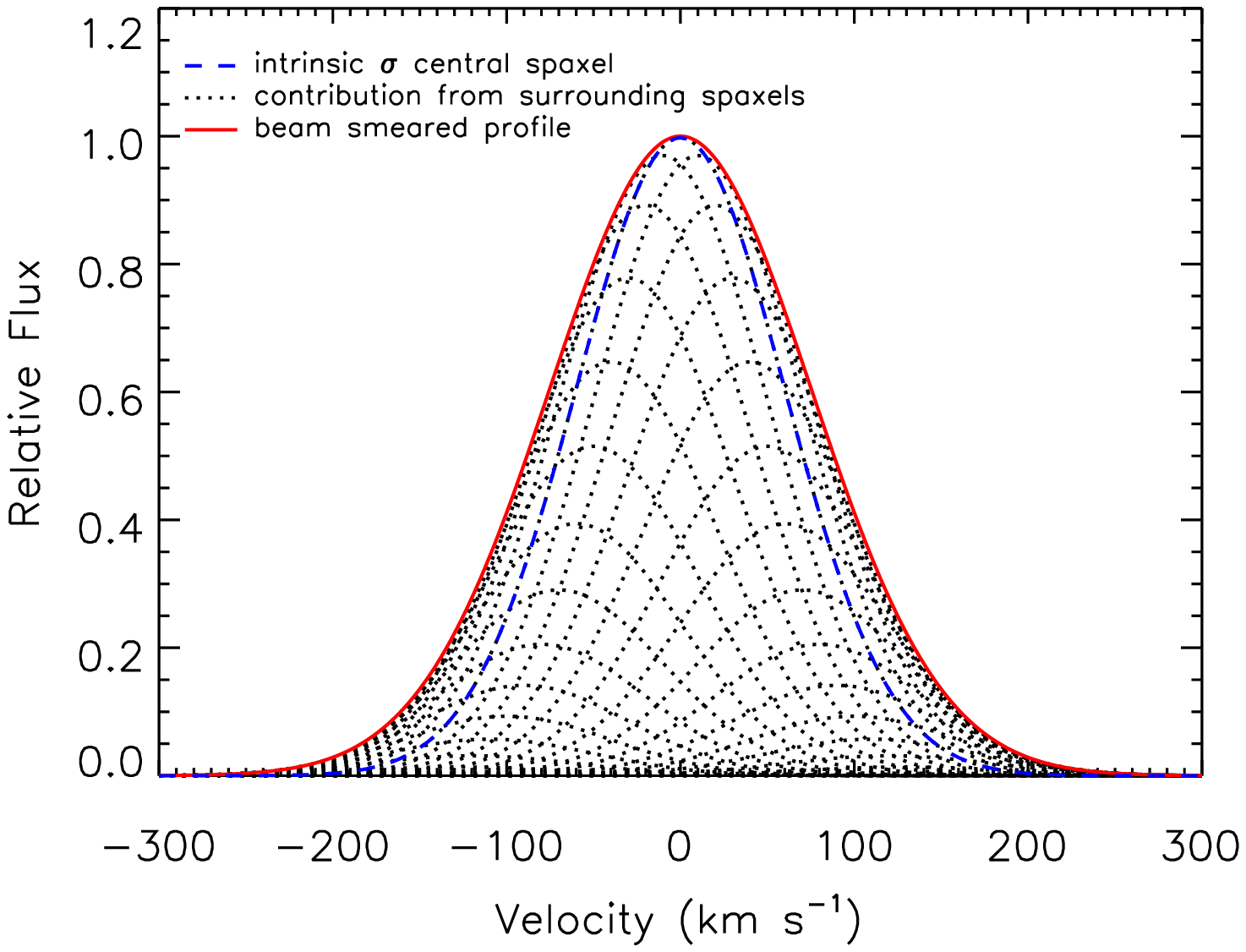}
\includegraphics[scale=0.5, trim=0 0 0 0, clip=true]{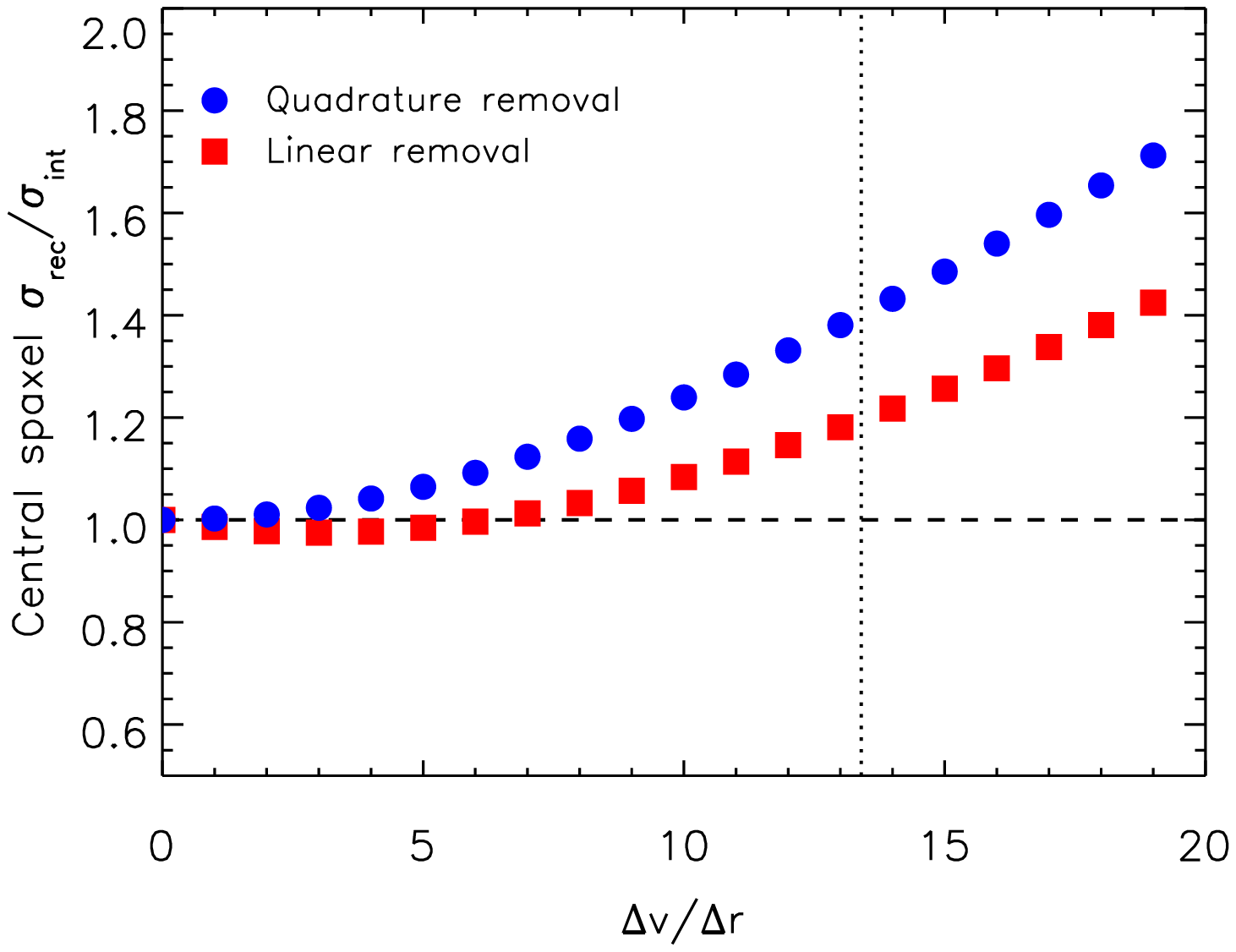}

  \caption[]{{\it Upper:} A 1-dimensional representation of the contributions to the velocity dispersion of the central spaxel from other spaxels due to beam smearing in an artificial galaxy. {\it Lower} The ratio of the recovered to intrinsic velocity dispersion of the central spaxel when correcting the beam smearing using the linear and quadrature techniques plotted as a function of $\frac{\Delta V}{\Delta R}$. This demonstrates that the linear removal is an improvement over the quadrature. The vertical dotted line at $\frac{\Delta V}{\Delta R} R=13.4\rm \, km\,s^{-1} spaxel^{-1} $ is the median maximum spaxel value of $\frac{\Delta V}{\Delta R}$ for our galaxies which shows that the linear removal results in only a $20\%$ residual compared to a $40\%$ excess when removed in quadrature.}
  
   \label{fig:beam}
\end{figure}

}

\end{document}